\documentclass{emulateapj}

\newcommand{\BP}{Ballesteros-Paredes}
\newcommand{\cwnm}{c_{\rm WNM}}
\newcommand{\Eg}{E_{\rm g}}
\newcommand{\Ek}{E_{\rm k}}
\newcommand{\Eth}{E_{\rm th}}
\newcommand{\eq}{{\rm eq}}
\newcommand{\gamef}{\gamma_{\rm e}}
\newcommand{\kms}{{\rm ~km~s}^{-1}}
\newcommand{\li}{\ell_{\rm inf}}
\newcommand{\Lbox}{L_{\rm box}}
\newcommand{\Lj}{L_{\rm J}}
\newcommand{\Mi}{{\cal M}_{\rm inf}}
\newcommand{\Mcl}{M_{\rm cl}}
\newcommand{\Mj}{M_{\rm J}}
\newcommand{\Msun} {M_\sun}
\newcommand{\Npart}{N_{\rm part}}
\newcommand{\Nss}{N_{\rm ss}}
\newcommand{\nthr}{n_{\rm thr}}
\newcommand{\pcc}{{\rm ~cm}^{-3}}
\newcommand{\psc}{{\rm ~cm}^{-2}}
\newcommand{\ri}{r_{\rm inf}}
\newcommand{\racc}{r_{\rm acc}}
\newcommand{\ti}{\tau_{\rm inf}}
\newcommand{\tff}{\tau_{\rm ff}}
\newcommand{\vi}{v_{\rm inf}}
\newcommand{\vrmsi}{v_{\rm rms,i}}
\newcommand{\VS}{V\'azquez-Semadeni}

\def\alamenos#1{$^{-#1}$}

\def\diezala#1{10$^{#1}$}

\slugcomment{Submitted to The Astrophysical Journal}

\shorttitle{Molecular Cloud Evolution II}
\shortauthors{V\'azquez-Semadeni et al.}

\begin{document}


\title{Molecular Cloud Evolution II. From cloud formation to the early
stages of star formation in decaying conditions}


\author{Enrique \VS\altaffilmark{1}, Gilberto C.\ G\'omez\altaffilmark{1}, A.\ Katharina
Jappsen\altaffilmark{2,3}, Javier \BP\altaffilmark{1}, Ricardo F.\ Gonz\'alez\altaffilmark{1} and Ralf
S.\ Klessen\altaffilmark{4}}

\altaffiltext{1}{Centro de Radioastronom\'ia y Astrof\'isica (CRyA),
Universidad Nacional Aut\'onoma de M\'exico,
Apdo. Postal 72-3 (Xangari), Morelia, Michoac\'an 58089, M\'exico}
\email{e.vazquez, g.gomez, j.ballesteros, rf.gonzalez@astrosmo.unam.mx}

\altaffiltext{2}{Astrophysikalisches Institut Potsdam, An der Sternwarte 16, 14482
Potsdam, Germany}

\altaffiltext{3}{Canadian Institute for Theoretical Astrophysics (CITA),
McLennan Physics Labs, 60 St. George Street, University of
Toronto, Toronto, ON M5S 3H860, Canada}

\altaffiltext{4}{Institut f\"ur Theoretische Astrophysik /
Zentrum f\"ur Astronomie der Universit\"at Heidelberg, Albert-\"Uberle-Str.
2, 69120 Heidelberg, Germany}


\begin{abstract}

We study the formation of giant dense cloud complexes and of stars
within them by means of SPH numerical simulations of the collision of
gas streams (``inflows'') in the warm neutral medium (WNM) at moderately
supersonic velocities. The collisions cause compression, cooling and
turbulence generation in the gas, forming a cloud that then becomes
self-gravitating and begins to collapse globally. Simultaneously, the
turbulent, nonlinear density fluctuations induce fast, local collapse
events. The simulations show that: a) The clouds are \emph{not} in a
state of equilibrium. Instead, they undergo secular evolution. During
the early stages of the evolution, their mass and gravitational energy
$\Eg$ increase steadily, while the turbulent energy $\Ek$ reaches a
plateau. b) When $\Eg$ becomes comparable to $\Ek$, global collapse
begins, causing a simultaneous increase in $|\Eg|$ and $\Ek$ that
maintains a near-equipartition condition $|\Eg| \sim 2 \Ek$. c) Longer
inflow durations delay the onset of global and local collapse, by
maintaining a higher turbulent velocity dispersion in the cloud over
longer times. d) The star formation rate is large from the beginning,
without any period of slow and accelerating star formation. e) The
column densities of the local star-forming clumps are very similar to
reported values of the column density required for molecule formation,
suggesting that locally molecular gas and star formation occur nearly
simultaneously. The MC formation mechanism discussed here naturally
explains the apparent ``virialized'' state of MCs and the ubiquitous
presence of HI halos around them. Also, within their assumptions, our
simulations support the scenario of rapid star formation
\emph{after} MCs are formed, although long ($\gtrsim$ 15 Myr)
accumulation periods do occur during which the clouds build up their
gravitational energy, and which are expected to be spent in the atomic
phase.

\end{abstract}




\section{Introduction} \label{sec:intro}

The evolution of molecular clouds (MCs) remains an unsolved
problem to date. While traditionally MCs have been thought of as
virialized structures in the interstellar medium (ISM) \citep[e.g., ][]{dBD80,
McKee_etal93, BW99, Mc99}, with relatively long lifetimes
\citep{BS80} and a significant delay before they begin forming
stars \citep[e.g.][]{PS00, PS02, TM04, MTK06}, recent studies have
suggested that MCs begin forming stars shortly after they themselves
form, and are non-equilibrium entities \citep[][ see also the reviews by
Mac Low \& Klessen 2004 and \BP\ et al.\ 2006]{BVS99, BHV99, Elm00,
HBB01, VBK03, PaperI, BH06, Ball06}. One approach that can shed light on this
problem is to perform numerical simulations of the MC formation process
by generic compressions in the warm neutral medium (WNM). Numerous
studies of this kind exist, although self-gravity has in general not
been included \citep[][ hereafter Paper I]{HP99, HP00, KI00, KI02, IK04,
AH05, Hetal05, Hetal06, PaperI}.

These studies have shown that a substantial
fraction of the internal turbulence, with velocity dispersions
of up to a few$\kms$, can be produced {in shock-bounded
layers} by a combined
dynamical+thermal instability, whose nature is still not well
determined \citep{Hunter_etal86, WF98, WF00, KI02,
IK04, Hetal05, Hetal06, FW06}. Paper I also showed that the dense
gas, there defined as the gas with densities 
larger than $100 \pcc$, is at systematically higher
thermal pressures than the mean interstellar value by factors of
1.5--5, implying that MC formation by colliding streams of diffuse
gas can account at least partially for the clouds' excess
pressure and turbulent nature. 

Only a few works have included self-gravity in simulations of
colliding flows. The pioneering work of
\citet{Hunter_etal86} considered the collision of supersonic
streams within MCs, finding that the
fragments produced by the dynamical instability (which they
termed ``Rayleigh-Taylor-like'') were able to accrete mass until
they became gravitationally unstable and collapse. However, this
study was designed to investigate flows \emph{within} MCs
rather than the formation of the clouds themselves,
and so it considered much smaller (sub-parsec) scales and
different density, temperature and cooling regimes than are
relevant for the process of MC formation. 

More recent numerical studies of MC formation not including
self-gravity \citep{KI02, Hetal05} have estimated that the
individual clumps formed in the compressed layer formed by the
colliding streams are themselves gravitationally stable,
although the entire compressed-gas complex would be
gravitationally unstable, had self-gravity been included.

This raises the issue of whether the clumps can be considered in
isolation.  The question of which gas parcels are involved in the
collapse of a particular object is a delicate one, and lies, for
example, at the heart of the debate of whether competitive
accretion is relevant or not in the process of star formation
within molecular clouds: A model in which clumps are isolated
dense objects immersed in a more tenuous, unbound medium leads to
the conclusion that competitive accretion is irrelevant
\citep{KMK05}. Instead, competitive accretion plays a central
role \citep{BBCP97, KBB98, BBCP01, KB00, KB01, K01a, K01b, BB06} in a scenario
in which the local density peaks are the ``tips of the iceberg''
of the density distribution, the flows are organized on much
larger scales
\citep{BVS99, BHV99, BBB03}, and the evolution and energetics of
the density peaks cannot be considered in isolation, but as part
of the global flow instead \citep{BVS99, KHM00, HMK01, SVB02, TP04, TP05, Ball06,
Dib_etal06}. In a similar fashion, then, accurate determination
of whether gravitational collapse can be triggered by
moderate-Mach number stream collisions in the WNM, 
requires to self-consistently include self-gravity in the
numerical simulations.

{Simulations of molecular cloud formation by the passage
of a self-gravitating, clumpy medium through Galactic spiral shocks
\citep{BDRP06, DBP06} have shown that realistic velocity dispersions and
densities can be generated in the resulting post-shock clouds, which are
furthermore driven to produce local collapse events leading to star
formation, although a large fraction of the mass in the clouds remains
gravitationally unbound. However, because these simulations were
isothermal, the gas had to be assumed to already be clumpy and at very
low temperatures ($\sim$ 100 K) \emph{previous} to the passage through the
shock, while recent studies including cooling leading to thermal
bistability of the atomic gas \citep[][ Paper I]{KI02, AH05, Hetal05, Hetal06} 
suggest that the clumpiness is generated by the shock compression, which
nonlinearly triggers the thermal instability, induces supersonic
turbulence, and causes the formation
of dense clumps in the compressed layers.}

It is {thus}  essential to model
the evolution of the cloud within the frame of its more diffuse
environment {including both self-gravity and cooling}. Not
only this is indispensible in order to model {the
generation of turbulence and clumpiness, but also to model the cloud's
evolution as it exchanges mass and energy with its surroundings. Note
that the generation of turbulence by the compression itself implies that
the turbulence within the cloud is driven while the large-scale
converging motions persist, and decaying afterwards. This is a mixed
regime that also requires self-consistent modeling of the cloud and its
surroundings, rather than, for example, random Fourier driving in closed
boxes.}

{Finally, such a unified (or ``holistic'') description of
molecular cloud formation and evolution, as outlined by \citet{VBK03},
including the star formation epoch, allows also the investigation of a
number of key problems, such as the evolutionary and star-formation
timescales, the evolution of the cloud's gravitational and kinetic
energies, and whether it settles into virial equilibrium or not, and the
effect of self-gravity on the cloud's physical conditions.}

In this paper we investigate
numerically the evolution of interstellar gas as it collects from the
WNM, {shocks, suffers a phase transition to a cold and
dense state}, and
finally begins to form stars, by means of numerical simulations of
colliding gas streams in the WNM, in the presence of self-gravity and
cooling leading to thermal instability. Since this is a problem that
naturally collects large amounts of gas in small volumes, first by
compression-triggered thermal instability
 and then by gravitational instability, it is
convenient to use some Lagrangian numerical scheme, and we opt for a
smoothed particle hydrodynamics code, sacrificing the possibility of
including magnetic fields. We also omit for now a self-consistent
description of the stellar feedback on the gas {and of the
chemistry}. Thus, our simulations will not
adequately describe the latest stages of MC evolution {(in
which stellar energy feedback may significantly influence the dynamics
of the cloud) nor are they capable of predicting the transition from
atomic to molecular gas accurately. We defer these tasks to subsequent
papers}.

The plan of the paper is as follows. In \S \ref{sec:quali} we first
give a qualitative description of the physical system, the main
phenomena involved, and our expectations for the evolution. In \S
\ref{sec:model} we describe the numerical code, the
physical physical setting, and the choice of parameters. Then, in
\ref{sec:results} we describe the results concerning the evolution of
the cloud's mass and its kinetic and gravitational energies, and the
onset of star formation. In \S \ref{sec:discussion} we discuss
the implications and limitations of the recent work. Finally, in \S
\ref{sec:concl} we present a summary and draw some conclusions.


\section{Qualitative discussion} \label{sec:quali}

It is convenient to first present an overview of the system, its
expected evolution, and the physical processes on which it relies
(see also the description in Paper I). Since
we are interested in studying the formation of a dense cloud out
of the diffuse medium, we take as initial condition a uniform-density
region within the WNM, of a few hundred parsecs across, with an
initially compressive velocity field consisting of two
oppositely-directed inflows with speed $\vi$ and Mach number $\Mi$, with
respect to the WNM temperature. This compressive field is
generically representative of either the general turbulence in the WNM,
or of motions triggered by some large scale {agent, such as
gravitational or Parker instabilities, or shell collisions}. Since the
velocity dispersion in the WNM is transonic 
\citep{KH87, HT03}, we consider values of $\Mi \sim 1$. 

As is well known, {the atomic
ISM is thermally bistable, with two stable phases being able to coexist
at the same pressure but different densities and temperatures, mediated
by a thermally unstable range \mbox{\citep{F65, FGH69}}}. The collision
of transonic WNM streams produces 
a thick shock-bounded slab, in which the gas is out of thermal
equilibrium between radiative heating and cooling. This shocked slab is
nonlinearly thermally unstable \citep{KI00, KN02}, and as it flows
towards the collision plane, it undergoes a phase transition to the cold
neutral medium (CNM), causing the formation of a thin cold layer,
\citep[] [Paper I]{HP99, HP00, KI00, KI02, IK04, AH05, Hetal05,
Hetal06}. Subsequently, the boundary of this layer is destabilized,
probably by a combination of thermal instability and nonlinear
thin-shell instability \citep{Vish94}, producing fully-developed 
turbulence in the layer for sufficiently large $\Mi$
\citep[][ Paper I]{Hunter_etal86, KI02, IK04, Hetal05, Hetal06}. As the
layer becomes turbulent, it thickens, and becomes a fully
three-dimensional structure (Paper I), to which we simply refer as ``the
cloud''. 

Moreover, the thermal pressure in the dense cloud is in pressure balance
with the total (thermal + ram) pressure in the inflows,
which is significantly larger than the mean thermal pressure in
the ISM (Paper I). Consequently, the density in the cold layer
can reach densities significantly larger than those typical of
the CNM. Thus, this cloud is turbulent, dense, and overpressured
with respect to the mean ISM pressure; i.e., it has properties
typical of molecular clouds (Paper I). The main missing link in
this scenario is whether the atomic gas can be readily converted
to molecular, although several studies suggest it is feasible
\citep{Bergin_etal04, GM06}. Since we do not incorporate any
chemistry nor radiative transfer in our code, we cannot follow
this transition self-consistently, and simply assume that once
the \emph{dense} gas has a sufficiently large column density (see
below), it is rapidly converted to the molecular phase.

In summary, the compression produces a turbulent cloud of dense
and cold gas. As it
becomes denser and colder, its Jeans mass decreases
substantially. Moreover, its own mass is increasing, so it
rapidly becomes much more massive than its own Jeans mass.
Also, as the cloud becomes more massive, its total
gravitational energy (in absolute value) $|\Eg|$ increases
substantially, and eventually overcomes even its turbulent
kinetic energy, and the cloud begins collapsing as a whole. Because it
contains many Jeans masses, local collapse events begin to occur at
the nonlinear density peaks produced by the turbulence,
{if gravity locally overcomes all available forms of
support. Because the density enhancements are nonlinear, the local
collapse events occur on shorter timescales than the global contraction.}

The magnetic field should also be considered. \citet{HP00} have
shown that the transition from diffuse to dense gas can occur
even in the presence of magnetic fields aligned with or oblique to the
direction of compression. Moreover, most evidence, both theoretical and
observational, points towards MCs and giant molecular complexes
being magnetically supercritical or marginally critical at least
\citep{Crut99, HBB01, Bourke_etal01, CHT03}. Since the
magnetically supercritical case is not qualitatively different
from the non-magnetic case, the above discussion is applicable to
supercritical clouds as well.

\citet{HBB01} have approximately quantified the above scenario,
estimating that the accumulation time from the WNM should be of
the order of 10--20 Myr, the accumulated gas must come from
distances up to a few hundred parsecs, and that at the end of the
accumulation period, the gas should be becoming molecular and
self-gravitating at roughly the same time, because the column
density for self-shielding and formation of molecular gas,
\begin{equation} \label{eq:N_self_shield}
N_{\rm ss} \sim 1-2 \times 10^{21} \psc,
\end{equation}
is very similar to that required for 
{rendering} the dense gas
gravitationally unstable,
\begin{equation} \label{eq:N_grav_unst}
N_{\rm gi} \sim 1.5 \times 10^{21} P_4^{1/2} \psc,
\end{equation}
where $P_4$ is the pressure expressed in units of $10^4$ K $\pcc$
\citep[see also][]{FC86}. That is, the gas is converted to the molecular
phase roughly simultaneously with the onset of gravitational
instability, explaining why non-star-forming MCs are rare, at 
least in the solar neighborhood \citep{HBB01, Hart03, BH06}. Note,
however, that this near equality is a coincidence \citep[e.g.,
][]{Elm85, Elm93a}, since $N_{\rm gi}$ depends on the ambient pressure, and
therefore non-star-forming molecular clouds may be more common in
higher-pressure environments,  {possibly explaining observations that up to
30\% of the GMCs appear to be starless in some external galaxies}
\citep{Blitz_etal06}. {On the other hand, this result may
be due in part to incompleteness effects. Indeed, recent Spitzer
observations have uncovered young stellar objects in Galactic molecular
cores previously believed to be starless \mbox{\citep[e.g.,
][]{Craspi_etal05, Rho_etal06}}}.

In the remainder of the paper, we {
proceed to} quantify this scenario by means of
numerical simulations designed for this purpose.


\section{The model} \label{sec:model}

\subsection{Parameters and numerical scheme} \label{sec:params}

\begin{figure}
\plotone{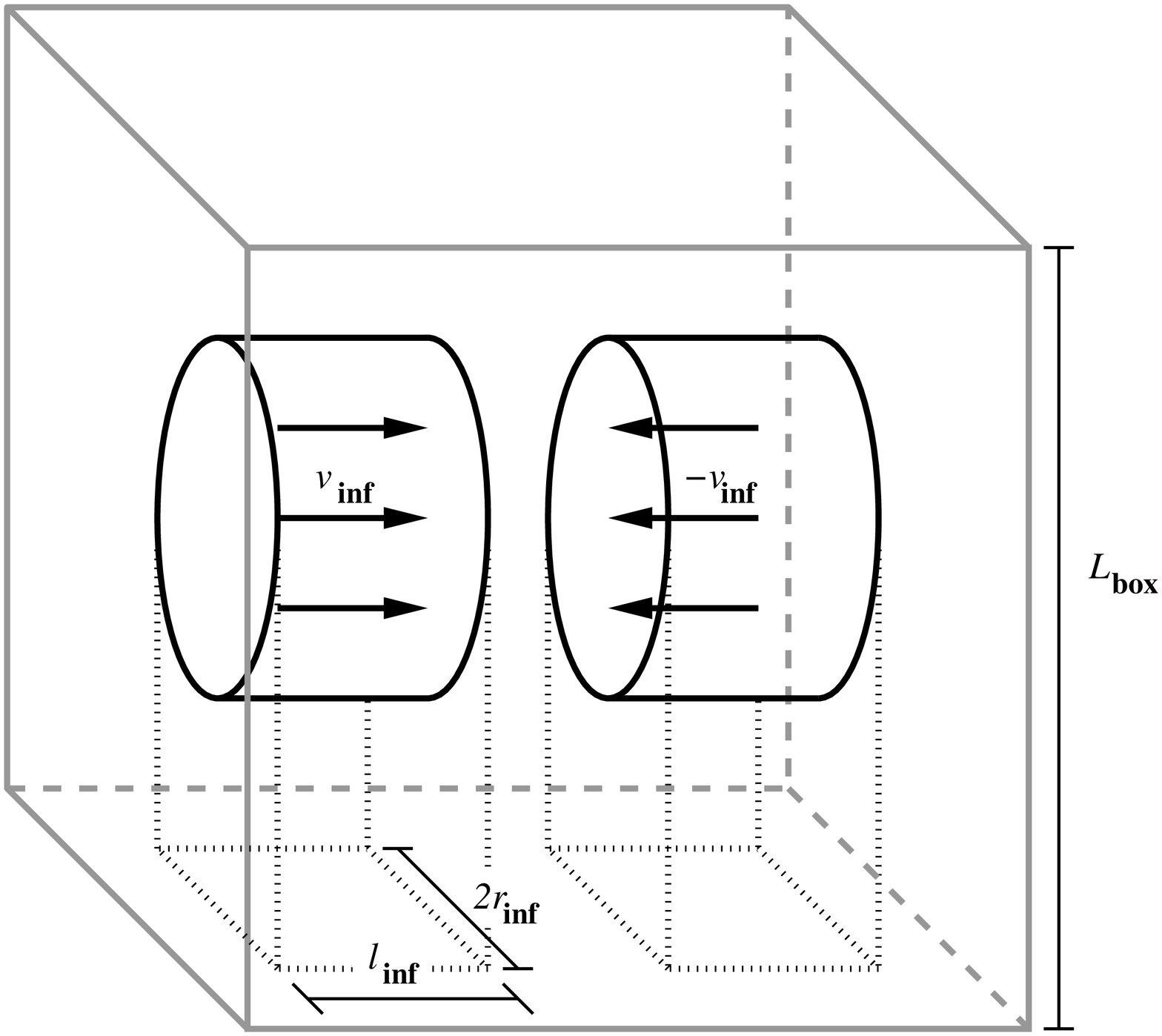}
\caption{
Setup of the simulations.
Within a cubic box measuring $\Lbox$ on a side, two cylindrical inflows,
each of length $\li$, radius $\ri$, and aligned with the $x$-axis
of the domain, are set to collide.
The inflows have a velocity corresponding to $\Mi = 1.22$ with
respect to the undisturbed medium.
Also, a fluctuating velocity field with amplitude $\vrmsi$ is added
at the begining of the run in order to trigger the dynamical
instability of the resulting compressed layer.
Table \ref{tab:run_params} lists the values used for these
parameters.
\label{fig:inflows}
}
\end{figure}

We consider a cubic box of length $\Lbox$ on a side, initially filled
with WNM at a uniform density of $n_0 = 1 \pcc$, with a mean molecular
weight of 1.27, so that the mean mass density is $\rho_0 = 2.12
\times 10^{-24}$ g cm$^{-3}$,
{which implies total masses of
$6.58 \times 10^4 M_\sun$ and $5.26 \times 10^5 M_\sun$
for the two box sizes used, $\Lbox = 128$ pc and $256$ pc,
respectively.
}
The gas has a temperature $T_0 = 5000$ K, the
thermal-equilibrium temperature at that density (cf.\ \S
\ref{sec:cooling}). Within the box, we set 
up two cylindrical and oppositely-directed inflows, each of length
$\li$, radius $\ri = 32$ pc, and speed $\vi=\pm 9.20~\kms$,
corresponding to a Mach number $\Mi = 1.22$ with respect to the sound
speed of the undisturbed WNM,
$\cwnm = 7.536 \kms$
{(see fig. \ref{fig:inflows}).
For the two values of $\li$ we consider {here} ($48$ pc
and $112$ pc), the 
masses of the cylindrical inflows
are $1.13 \times 10^4~M_\sun$ and $2.64 \times 10^4~M_\sun$, respectively.
}
To trigger the dynamical
instability of the compressed slab, a fluctuating velocity field
(computed in Fourier space) is initially imposed throughout the box at
wavenumbers $4 \le k\Lbox \le 8$, and having amplitude $\vrmsi$. The
streams are directed along the $x$ direction and collide at half the $x$
extension of the box. Table \ref{tab:run_params} gives the values of
these parameters for the simulations presented here. The runs are named
mnemonically as L{\it xxx}$\Delta v$0.{\it yy}, where xxx indicates the
physical box size in 
parsecs and 0.{\it yy} indicates the value of $\vrmsi$ in $\kms$.

We use the N-body+smoothed particle hydrodynamics (SPH) code GADGET
\citep{SYW01}, as modified by \citet{Jappsen_etal05} to include sink
particles, a prescription to describe collapsed objects without
following their internal structure.  We use $1.64 \times 10^6$
particles for the runs with $\Lbox =128$ pc, and $3.24 \times 10^6$
particles for the run with $\Lbox = 256$ pc.
Since the larger-box
run contains 8 times more mass than the smaller-box ones
{but uses only twice as many particles,}
{the minimum resolved mass in the larger-box run is
four times larger than in the smaller-box run.}
Unfortunately, the computing
resources available in our cluster did not allow us to further
increase the number of particles.

We take as units for the code a mass of $1 M_\sun$, a length $\ell_0 =
1$ pc and a velocity $v_0 = 7.362 \kms$, implying a time unit $t_0 =
l_0/v_0 = 0.133$ Myr.



\subsection{Sink particles} \label{sec:sinks}

Sink particles are created when a group of SPH particles become
involved in a local collapse event. At that point, the SPH
particles are replaced by a sink particle whose internal
structure is not resolved anymore. The sink particle inherits the
total angular momentum and mass of the SPH particles it replaces,
and henceforth moves only in response to its inertia and the
gravitational force, although it can continue to accrete SPH
particles, should they come sufficiently close to the sink
particle.

The creation of a sink particle requires a number of conditions
to be satisfied. First, the local density should exceed some
user-defined threshold value $\nthr$. If this occurs, then the
code computes the total mass and angular momentum of the group of
particles above $\nthr$ to determine whether they are
gravitationally bound. If they are, then they are collectively
replaced by the sink. At subsequent times, further SPH particles
can be accreted by the sink if they come within an ``accretion
radius'' $\racc$ of the sink and are gravitationally bound to
it. In all cases we use $\nthr = 10^5
\pcc$ and $\racc = 0.04$ pc. {This value of $\racc$ amounts to roughly
1/3 of the Jeans length at $\nthr$}. Due to the limited mass resolution
we can afford, the sinks should in general be considered star clusters
rather than isolated stars, although some of them may actually
correspond to single stars. Nevertheless, with this caveat in mind, we
refer to the sinks as ``stars'' in a generic way.


\subsection{Heating and cooling} \label{sec:cooling}

The gas is subject to heating and cooling functions of the form  

\begin{eqnarray} 
\Gamma &=& 2.0 \times 10^{-26} \hbox{ erg s}^{-1}\label{eq:heating}\\
\frac{\Lambda(T)}{\Gamma} &=& 10^7 \exp\left(\frac{-1.184 \times
10^5}{T+1000}\right) \nonumber \\
&+& 1.4 \times 10^{-2} \sqrt{T} \exp\left(\frac{-92}{T}\right)
~{\rm cm}^3.\label{eq:cooling}
\end{eqnarray}
These functions are fits to the various heating ($\Gamma$) and
cooling ($\Lambda$)  processes considered by
\citet{KI00}, as given by equation (4) of \citet{KI02}.\footnote{Note 
that eq.\ (4) in \citet{KI02} contains two typographical
errors. The form used here incorporates the necesary corrections,
kindly provided by H.\ Koyama.} The resulting thermal-equilibrium
pressure $P_{\rm eq}$, defined by the condition $n^2 \Lambda =
n \Gamma$ is shown in fig.\ \ref{fig:P_vs_rho_KI} as a function of
density. We have abandoned the piecewise power-law fit we have
used in all of our previous papers \citep[e.g.,][]{VPP95, PVP95, VPP96, VGS00,
GVSS01, GVK05}, including Paper I, mainly
because the present fit is applicable at densities typical of
molecular gas.

The usual procedure to apply cooling to the hydrodynamic
evolution involves the consideration of the cooling rate in the
Courant condition to restrict the simulation time step.  For the
problem at hand, the high densities reached behind the shocks,
for example, would imply an exceedingly small time step and the
simulation would become computationally unfeasible.  But this
appears completely unnecessary, because all this means is that
the thermal evolution happens faster than the dynamical one,
and therefore, as far as the hydrodynamics is concerned,
it is instantaneous.  Therefore, it is more convenient to use an
approximation to the thermal evolution of the gas, which
constitutes an extension of that used in Paper I, and which
allows us to simply correct the internal energy after the
hydrodynamic step has been performed, with no need to adjust the
timestep. 

Consider the thermal equilibrium temperature, $T_{\eq}$, as a function
of the gas density, and the corresponding internal energy density
$e_{\eq}$ ($T_\eq$ is a well defined function of the density $n$
as long as  $\Lambda(T)$ is monotonic. This is true within the
range of temperatures occurring in the simulations).
Consider also the time required to radiate the thermal energy excess
(if $T > T_{\eq}$; acquire the energy deficit, if $T < T_{\eq}$)

\begin{equation} \label{eq:tau_lam}
\tau_\Lambda = \left| \frac{e-e_{\eq}}{n^2 \Lambda - n \Gamma}\right|.
\label{eq:tau_Lambda}
\end{equation}

\noindent
Then, we compute the new internal energy density $e'$, after a
time step $dt$, as

\begin{equation} \label{eq:new_e}
e' = e_\eq + ( e - e_\eq ) \exp( -dt / \tau_\Lambda).
\end{equation}

\noindent
Note that, if the gas is cooling down (or heating up) rapidly,
$\tau_\Lambda \ll dt$, $\exp(-dt/\tau_\Lambda) \rightarrow 0$ and
the gas immediately reaches its equilibrium temperature, without
ever overshooting beyond that value.  Conversely, if the gas is
at very low density or is close to the equilibrium temperature,
$\tau_\Lambda \gg dt$, and equation (\ref{eq:new_e}) becomes,

\begin{equation} \label{eq:small_dt}
e' = e - dt ( n^2 \Lambda - n \Gamma ),
\end{equation}

\noindent
where we, again, use the fact that $\Lambda(T)$, as given by equation
(\ref{eq:cooling}), is a monotonic function of the temperature.


\subsection{Gravitational instability in cooling media}
\label{sec:grav_cool}

\begin{figure}
\plotone{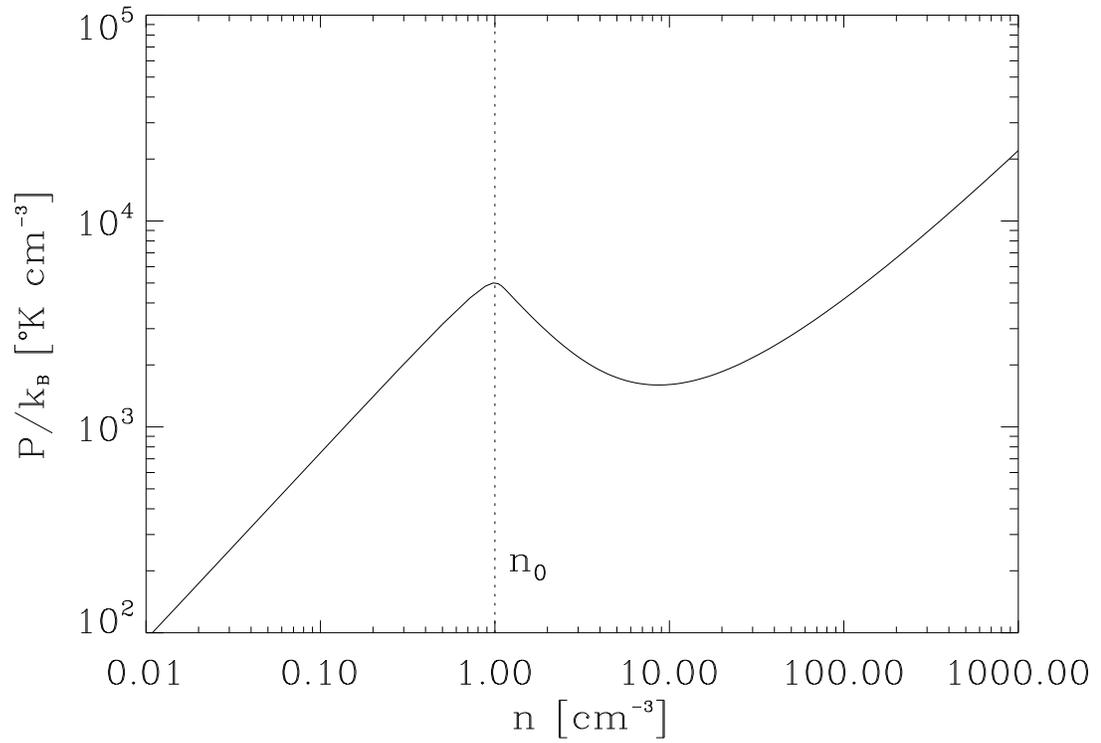}
\caption{Thermal-equilibrium pressure \emph{versus} density for the
cooling and heating functions given by eqs.\ (\ref{eq:heating}) and
(\ref{eq:cooling}). The vertical \emph{dotted} line indicates the
uniform initial density in the simulations. The tangent to the curve at
that location has a near-zero slope, indicating $\gamef \approx
0$. At $n = 0.2 \pcc$, $\gamef = 0.83$. \label{fig:P_vs_rho_KI}
}
\end{figure}

The standard Jeans instability analysis is modified in a medium
in which the balance between heating and
{\emph{instantaneous}}
cooling produces a net
polytropic behavior, characterized by an effective polytropic exponent
$\gamef$, which is the slope of the thermal-equilibrium pressure
\emph{versus} density curve (fig.\ \ref{fig:P_vs_rho_KI}). In this case,
the Jeans length is found to be given 
{by} \citep[e.g., ][]{Elm91, VPP96}
\begin{equation}
\Lj \equiv \left[\frac{\gamef \pi c^2}{\gamma G \rho_0}\right]^{-1/2},
\end{equation}
where $c$ is the adiabatic sound speed and $\gamma$ is the heat capacity
ratio of the gas. Equivalently, this instability criterion can be
described as if the effective sound speed in the system were given by
$c_{\rm eff} = \sqrt{\gamef k T/\mu}$, where $k$ is Boltzmann's constant
and $\mu$ is the mean molecular weight.
{Note that, when $\gamef=0$, the thermal pressure does not react to
density changes and therefore is unable to oppose the 
collapse,  {rendering} the medium gravitationally unstable at all scales
($\Lj=0$).
In a more realistic situation where the cooling is not instantaneous,
the Jeans length will be limited by the cooling length.
In this work,} {the cooling has a finite characteristic
timescale, given by eq.\ (\ref{eq:tau_Lambda}).}

At the initial uniform density $n_0= 1 \pcc$ and temperature $T_0 =
5000$ K, $\gamef \sim 0$ and the Jeans length is also near
zero. This can be seen in fig.\ \ref{fig:P_vs_rho_KI}, where the
vertical dotted line shows the initial uniform density, at which the
local tangent to the $P_{\rm eq}$ \emph{vs.} $\rho$ curve is seen to have a
nearly zero slope. However, when the compression produces a dense cloud at the
midplane of the box, the rest of the gas decreases its density and
enters the fully stable warm phase. For reference, at a density $n = 0.5
\pcc$ and a corresponding equilibrium temperature $T_{\rm eq} =6300$ K,
$\gamef = 
0.834$ and $\Lj = 1324$ pc, implying a Jeans mass $\Mj = 3.47 \times
10^7 M_\sun$. Under these conditions, our runs with $\Lbox = 128$ and
$\Lbox =256$ pc would initially contain $\sim 10^{-3}$ and $\sim
10^{-2}$ Jeans masses, respectively.


\section{Results} \label{sec:results}

\subsection{Mass evolution} \label{sec:mass} 

\begin{figure}
\epsscale{1.1}
\plottwo{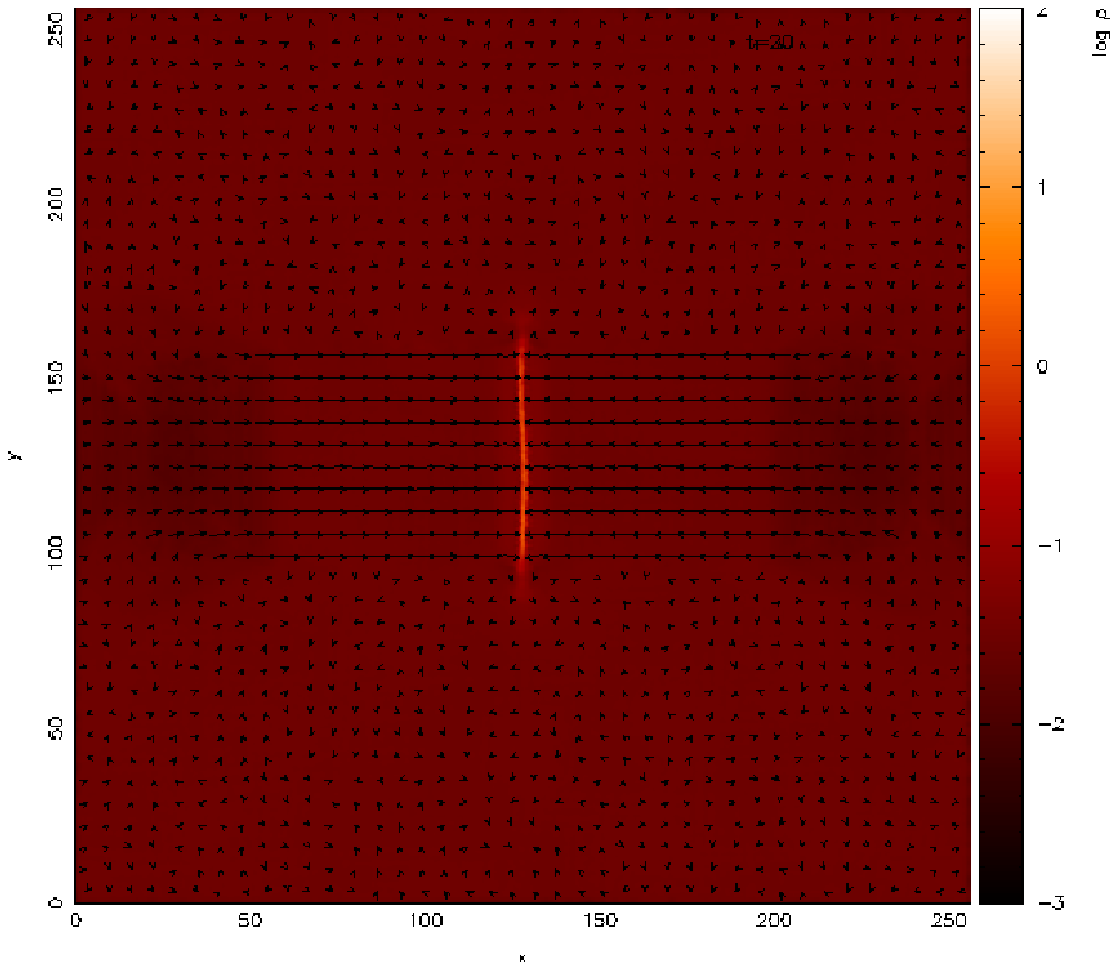}{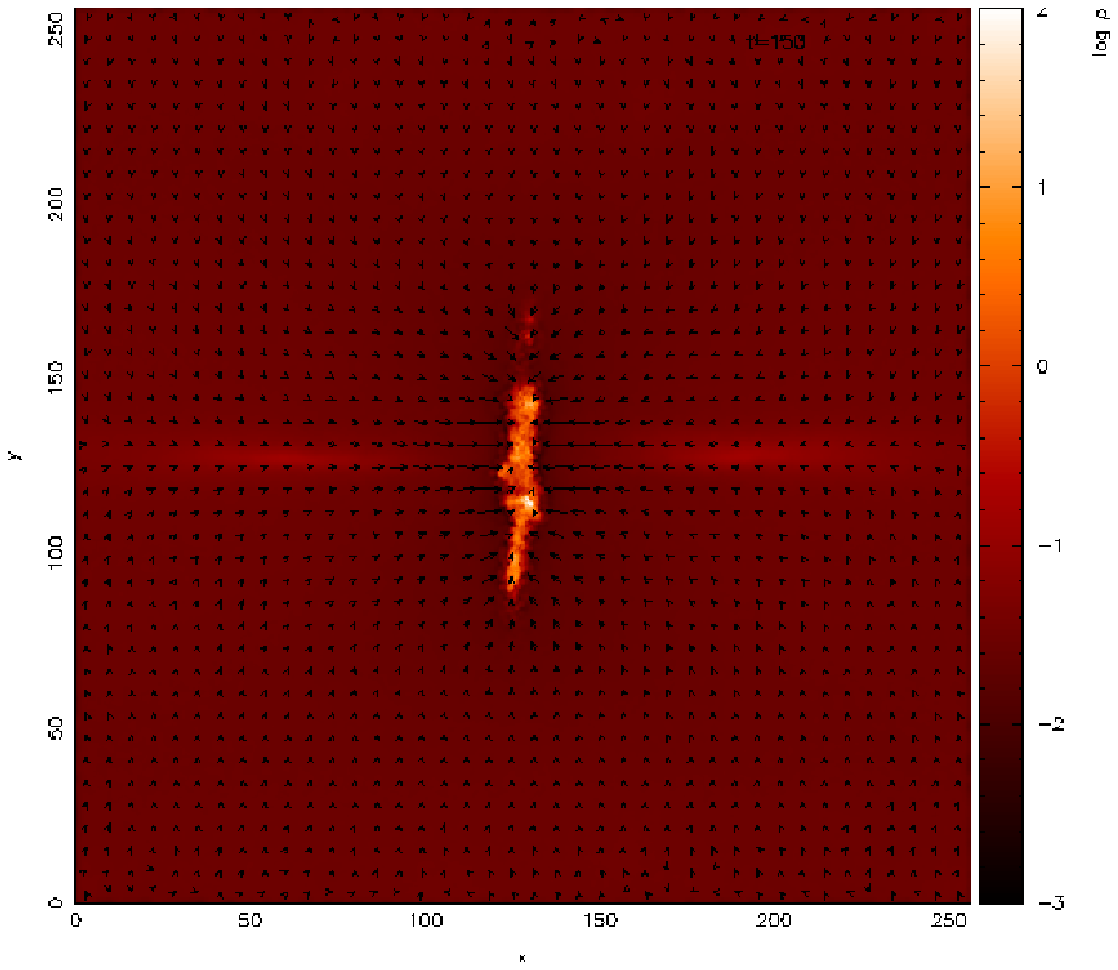}
\caption{Edge-on views of run L256$\Delta v$0.17 at $t=2.66$ Myr
(\emph{left}) and 20.0 Myr (\emph{right}). The arrows indicate the
velocity field. This figure is available as an mpeg animation in the
electronic edition of The Astrophysical Journal. \label{fig:anim_256_edgeon}
}
\end{figure}

\begin{figure}
\epsscale{1.1}
\plottwo{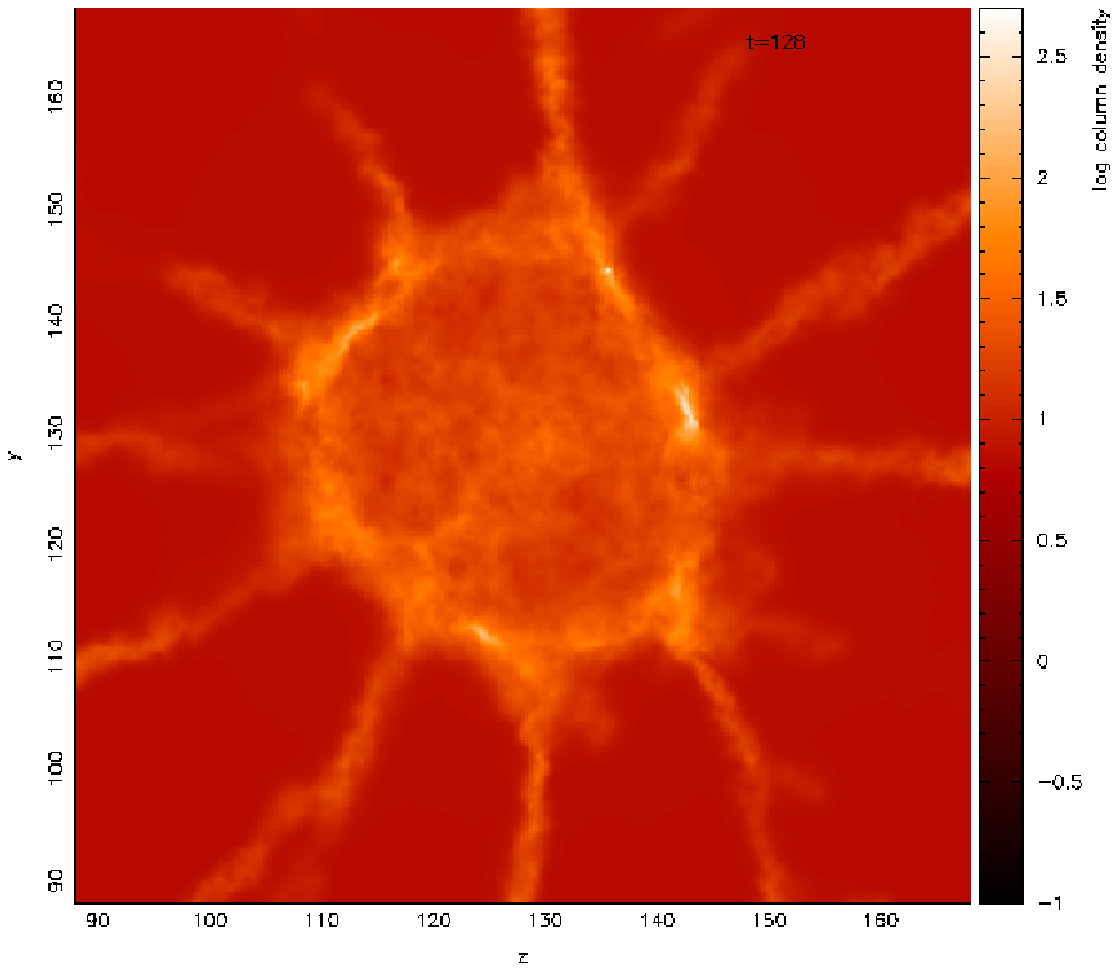}{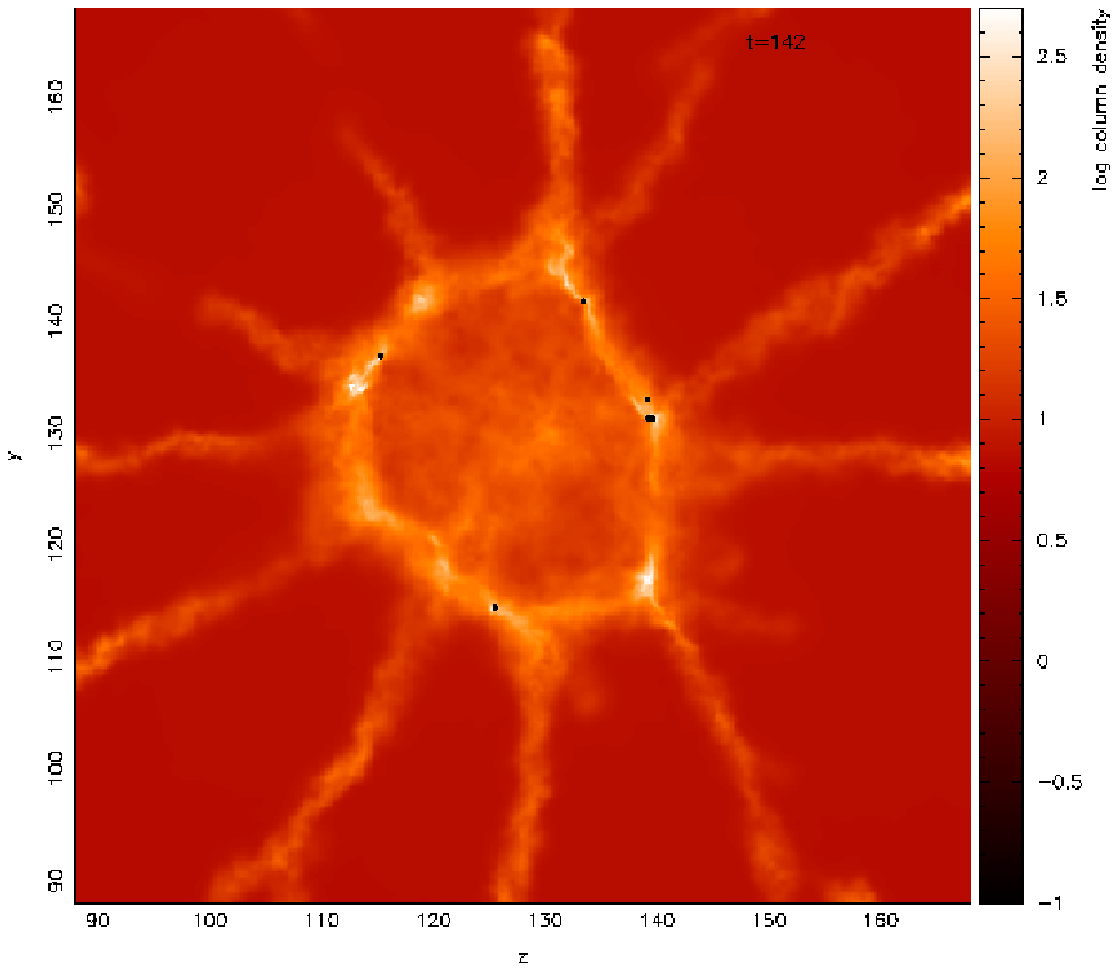}
\caption{Face-on views of run L256$\Delta v$0.17 at $t=17.0$ Myr
(\emph{left}) and 18.9 Myr (\emph{right}). The dots indicate the
projected positions of the sink particles This figure is available as an
mpeg animation in the electronic edition of The Astrophysical
Journal.. \label{fig:anim_256_faceon} }
\end{figure}

In general, the simulations evolve as described in Paper I and in
sec.\ \S \ref{sec:quali}. As an illustration, figs.\
\ref{fig:anim_256_edgeon} and \ref{fig:anim_256_faceon} show
selected snapshots of run L256$\Delta v$0.17 viewed edge-on and face-on,
respectively. In the electronic version of the paper, these
figures are also available as full-length animations. The times
indicated in the frames are in the code's internal unit, $t_0 =
0.133$ Myr. 

The animation shows that the gas at the collision site begins to
undergo a phase transition to the cold phase (the CNM) at roughly
one cooling time after the collision \citep{HP99, PaperI},
becoming much denser and colder. Because this gas is in
equilibrium with the total (thermal+ram) pressure of the
inflow, the density overshoots far beyond that of standard CNM,
well into the realm of molecular gas, with mean densities of
several hundred $\pcc$ and temperatures of a few tens of
Kelvins. This implies a much lower Jeans mass than that in the
conditions of the initial WNM and quickly the cloud's mass
exceeds its Jeans mass.

{Defining the ``cloud'' as the material with $n > 50 \pcc$,
}
the cloud's mass becomes larger than its mean Jeans mass at $t \sim 2.2$
Myr in runs with $\Lbox = 128$ pc and at $t \sim 3$ Myr in the
run with $\Lbox = 256$ pc, at which times the mean density in the
dense gas is $\langle n \rangle \sim 65 \pcc$, the mean temperature is
$\langle T \rangle \sim 50$ K, the Jeans mass is $\Mj \sim 1200~M_\sun$,
and the free-fall time is $\tff \equiv
\Lj/c \sim 11$ Myr, where $c$ is the local sound speed. For
comparison, the total mass in the two cylinders,
{which contain the material that
{initially} builds up the cloud,
}
is $1.13 \times 10^4~M_\sun$ in the runs with $\Lbox = 128$ pc,
and $2.26 \times 10^4~M_\sun$ in the run with $\Lbox = 256$ pc.
Thus, the cloud
eventually becomes much more massive than its mean Jeans mass.
For example,
{in run L256$\Delta v$0.17
}
at $t=17.26$ Myr, the (mass-weighted) mean
density of the dense gas is $\langle n \rangle \sim 600 \pcc$, the mean
temperature is $\langle T\rangle \sim 37$ K, and $\gamef \sim 0.75$. These
values imply a mean effective Jeans mass $\Mj \sim 300~M_\sun$, while at that
time the cloud contains $\sim 2.35 \times 10^4~M_\sun$.
{Note that the cloud{+sink} mass might exceed the mass in the
inflows, 
since ambient material surrounding the cylinder is dragged along
with the flow, effectively increasing the amount of gas in the
inflow} {(see below)}.

The animations show that global collapse does indeed occur, although
{at significantly later times than when the cloud's
mass becomes larger than the Jeans mass},
probably reflecting the role of turbulence as
an opposing agent to the collapse. Indeed, the clouds in the 128-pc
boxes begin global 
contraction at $t \sim 8$ Myr, while the cloud in the 256-pc box, with
more than twice the inflow duration, begins contracting 
at $t \sim 12.5$ Myr. 

Figure \ref{fig:multi_evo} (\emph{top} panel) shows the evolution
of the {cloud's mass}
in run L256$\Delta v$0.17, together with the mass in
collapsed objects (sinks), to be discussed further in \S \ref
{sec:energy_sf}. It is clearly seen that the \emph{cloud's mass is not
constant, but rather evolves in time}. Gas at 
{cloud} densities
first appears after 2 Myr of evolution, and the cloud's mass
increases monotonically until $t \sim 19.5$ Myr, {at which
time it reaches a maximum value $\Mcl \sim 2.7 \times 10^4
\Msun$}. After this time, 
it begins to decrease because of the rapid conversion of gas to
stars. 

By the
end of the simulation, the total mass in stars plus dense gas is
almost three times the initial mass in the inflows. In practice,
the inflows last longer than the simple estimate $\ti \equiv \li/\vi$ would
indicate, and involve more mass than the mass initially within
the cylinders. This is due to the fact that, as the cylinders
begin advancing, they leave large voids behind them that have the
double effect of slowing down the tails of the inflows and of
dragging the surrounding gas behind the cylinders. For
$\vi = 9.20 \kms$, $\ti= 5.22$ Myr for runs with $\li =48$ pc
(L128$\Delta v$0.24 and L128$\Delta v$0.66), and $\ti = 12.17$ Myr for
the run with 
$\li = 112$ pc (L256$\Delta v$0.17). Instead, in the first two runs, the inflow
actually lasts $\sim 11$ Myr, while in the latter run it lasts
$\sim 22.5$ Myr. This can be observed in the animation of fig.\
\ref{fig:anim_256_edgeon}, in which the velocity field arrows
clearly show the gradual decay of the inflows, and their longer
duration compared to the simple linear estimate.


\subsection{Cloud disruption} \label{sec:cloud_disrup}

Our simulations do not include any feedback from the stellar objects,
while this process is probably essential for the energy balance and
possibly the destruction of the cloud
\citep[e.g., ][]{Cox83, Cox85, FST94, HBB01, Matz02, NL05, LN06,
Mell_etal06}. Thus, the 
simulations cannot be safely considered realistic after  {their} stellar content
would likely disrupt the cloud. We now present an \emph{a posteriori}
estimation of the time at which this occurs.

%

\citet{FST94} have suggested that the maximum number
of OB stars that the cloud can support at any time is given by the
number of H~II regions required to completely ionize the cloud.
In particular, they found that the maximum number of massive stars
that can be formed within a molecular cloud is

\begin{equation}
N_{\rm OB, inside} \simeq {16 M_{c,4}\ n_3^{3/7}\over F_{48}^{5/7}(c_{s,15}\
                   t_{\rm MS,7})^{6/7}  }  ,
\label{eq:franco4insidestars}
\end{equation}
where $M_{c,4}$ is the mass of the cloud, in units of \diezala 4
$\Msun$, $n_3$ is the number density in units of \diezala 3
cm\alamenos 3, $F_{48}$ is the photoionizing flux from the massive stars,
in units of \diezala{48} s\alamenos 1, $c_{s,15}$ is the isothermal
sound speed in units of 15 $\kms$, and $t_{\rm MS,7}$ is a characteristic OB
star main sequence lifetime, in units of \diezala 7 yr.  However, as
\citet{FST94} pointed out, clouds are more efficiently
destroyed by stars at the cloud edge because the lower external
pressure ensures that the ionized gas expands rapidly away from
the cloud, driving a fast ionization front into the dense
material.  Thus, in this case, these authors find that the
maximum number of OB stars that the molecular cloud can support
is given by:

\begin{equation}
N_{\rm OB, edge} \simeq {3 M_{c,4}\ n_3^{1/5}\over
                   F_{48}^{3/5}(c_{s,15}\ t_{\rm MS,7})^{6/5} } .
\label{eq:franco4edgestars}
\end{equation}

In order to determine the time at which the MC in our simulation
has formed a sufficient number of stars to disrupt it, we proceed
as follows.  Since the individual sink particles in the
simulations in general correspond to clusters rather than to
single stars, we cannot use the sink particle masses
directly. Instead, for each temporal output of our simualtion, we
fit a standard initial mass function \citep{Krou01} to the total
mass in sink particles at that time, to obtain the the ``real''
distribution of stellar masses in the simulation at that
time.\footnote{Note that this estimate for the number
of massive stars as a function of time is probably a lower limit,
as recent simulations suggest that massive stars in clusters tend
to form first, while stars formed later tend to be less massive
because of the competition for accretion with the existing
massive stars \citep[e.g.][]{Bon05}.} We then use Table 1 of
\citet{DFS98}, which gives the 
ionizing flux as a function of stellar mass, to estimate the flux
associated to each mass bin, and then integrate over all relevant
masses to obtain the total $F_{48}$. 

Now, from eqs.\ (\ref{eq:franco4insidestars}) 
{or} (\ref{eq:franco4edgestars}) one can solve for $M_{c,4}$ 
as a function of $N_{\rm OB}$, to find the minimum
cloud mass that survives the ionizing radiation from the
existing OB stars in the simulation. When this minimum surviving
mass is larger than the actual cloud mass, we expect that the
cloud will be disrupted. Since we define the
``cloud'' as the gas with  
densities above 50$\pcc$, we take $n_3 = 0.05$. Also, for simplicity
we assume $c_{s,15}\ t_{\rm MS,7} \sim 1$.

\begin{figure}
\epsscale{1.}
\plotone{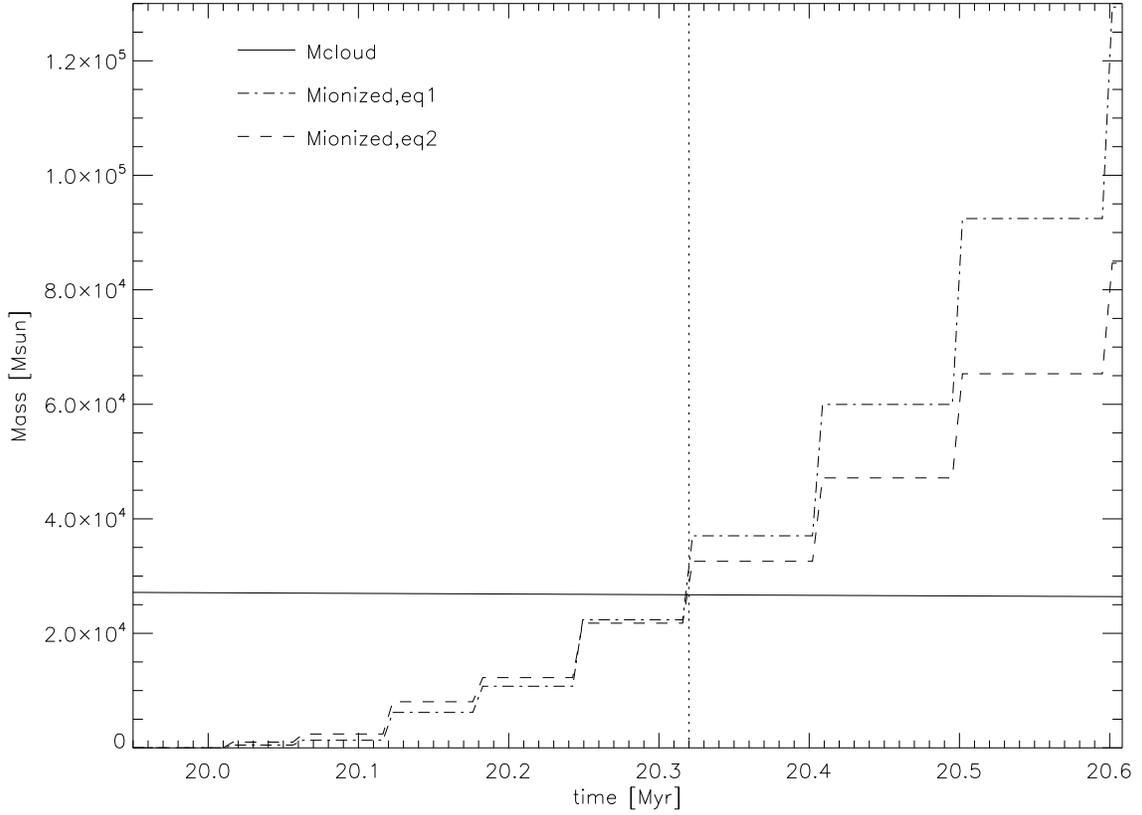}
\caption{Evolution of the cloud mass and of the minimum mass that
is not dispersed by stellar ionizing radiation under the two
estimates given by eqs.\ (\ref{eq:franco4insidestars}) and
(\ref{eq:franco4edgestars}) for run L256$\Delta v$0.17. The cloud is
expected to be 
disrupted when the minimum cloud mass for non-disruption exceeds the
actual cloud mass, at the time indicated by the vertical \emph{dotted}
line. \label{fig:cloud_disrup} 
} 
\end{figure}

For run L256$\Delta v$0.17, fig.\ \ref{fig:cloud_disrup} shows the
evolution of the cloud's mass together with the minimum disrupted
cloud mass under the two estimates, eqs.\
(\ref{eq:franco4insidestars}) and (\ref{eq:franco4edgestars}). We
see that the cloud is expected to be disrupted at $t=20.3$ with
either estimate. 

At this time the mass in sinks is $M_{\rm sinks}
= 5.25 \times 10^3 M_\sun$ and the cloud's mass is $M_{\rm cl} =
2.67 \times 10^4 M_\sun$, so that the star formation efficiency (SFE)
up to this time has been
\begin{equation}
{\rm SFE} = \frac{M_*}{M_{\rm cl} + M_*} \approx 16\%.
\label{eq:SFE_disrup}
\end{equation}
This number is still larger than observational estimates for
cloud complexes \citep{Myers_etal86}, and suggests that
additional physical processes, such as longer inflow durations
(\S \ref{sec:eff_infl_sigma}) or magnetic fields (\S
\ref{sec:caveats}) may be necessary to further reduce the SFE.

We conclude from this section that after $t \approx 20.3$ Myr run
L256$\Delta v$0.17 is probably not realistic anymore, as far as the
evolution of a real MC is concerned, and we thus 
restrict most of our subsequent discussions to times earlier than that,
except when later times may be illustrative of some physical
process. Self-consistent inclusion of stellar feedback in the
simulations, similarly to the studies of \citet{NL05} and
\citet{LN06}, to investigate the final stages of MC evolution will
be the subject of future work.


\subsection{Turbulence evolution} \label{sec:turb} 

The collision also generates turbulence in the dense
gas, and so the cloud can be considered to be in a driven
turbulent regime while the inflows persist, and in a decaying
regime after the inflows subside, although the transition from
driven to decaying is smooth, rather than abrupt, since the
inflows subside gradually, as explained above. Figure
\ref{fig:multi_evo} (\emph{bottom} panel) shows the evolution of the velocity
dispersion $\sigma$ for the dense gas along each of the three
coordinate axes for run L256$\Delta v$0.17. During the first
10 Myr of the cloud's existence ($2 \lesssim t \lesssim 12$ Myr)
$\sigma_y \approx \sigma_z \sim 4
\kms$, although it is interesting to
note that the turbulence appears to be anisotropic, with $\sigma_x$
first increasing and then decreasing. It
reaches a maximum of more than twice as large ($\sim 8.5 \kms$) as that of
$\sigma_y$ and $\sigma_z$ at $t \approx 6$ Myr, reflecting the fact that
the inflows are directed along this direction. This suggests that the
generation of transverse turbulent motions is not 100\% efficient. Of
course, this effect is probably exaggerated by our choice of perfectly
anti-parallel colliding streams. Clouds formed by obliquely colliding
streams are likely to have more isotropic levels of turbulence.

After $t \sim 6$ Myr, $\sigma_x$ begins to decrease, reflecting the
weakening of the inflows, settling at $\sigma_x \sim 6 \kms$ by $t \sim
12$ Myr. At this time, $\sigma_y$ and $\sigma_z$ begin to
increase, reflecting the global collapse of the cloud on the $yz$ plane,
while $\sigma_x$ remains nearly stationary, unaffected by the global
planar collapse, until $t \sim 18$ Myr, at
which time it also begins to increase. This coincides with the onset of
star formation, and probably reflects the local, small-scale, isotropic
collapse events forming individual collapsed objects.


\subsection{Evolution of the density, pressure and temperature
distributions} \label{sec:hists} 

\begin{figure}
\plottwo{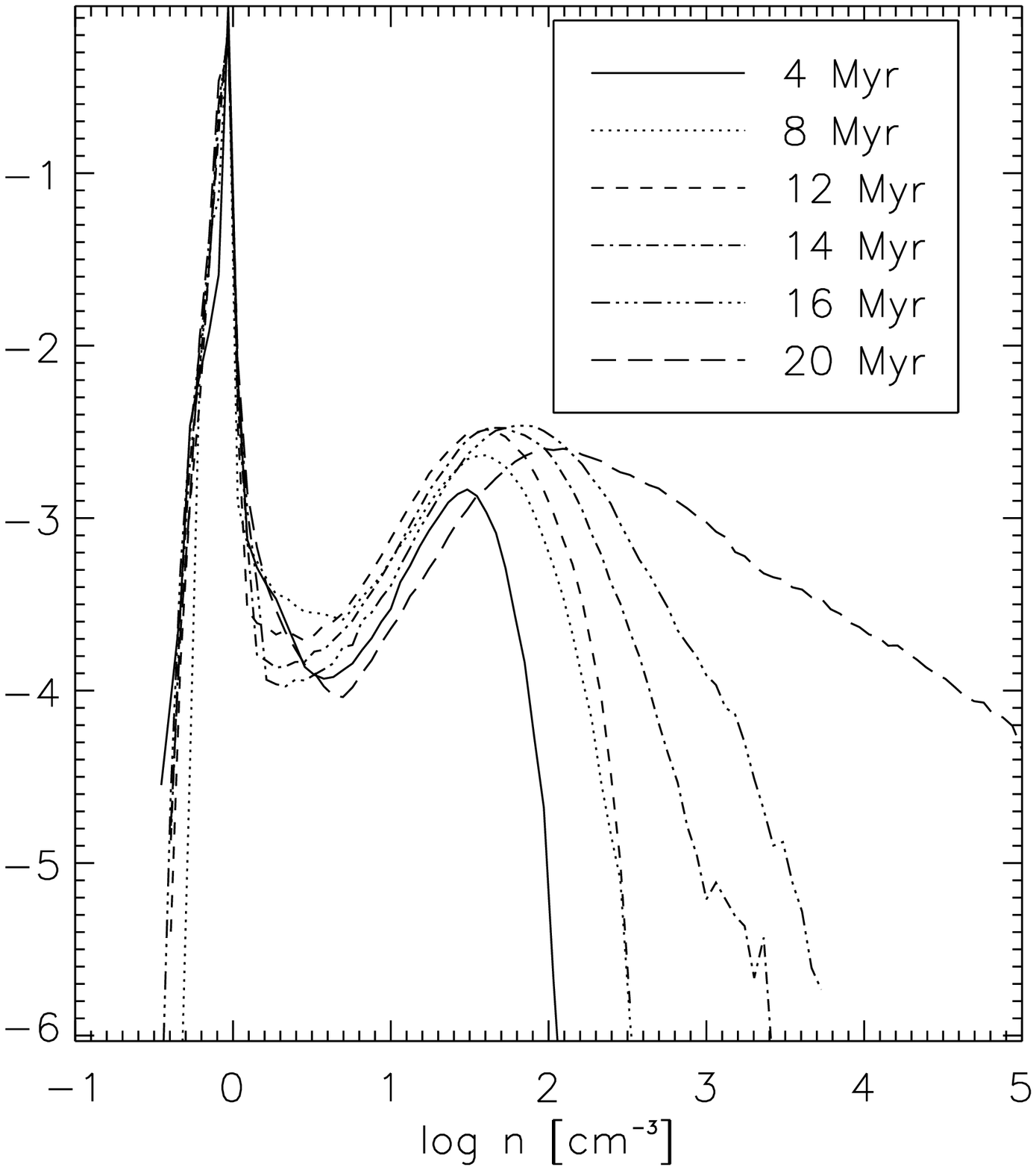}{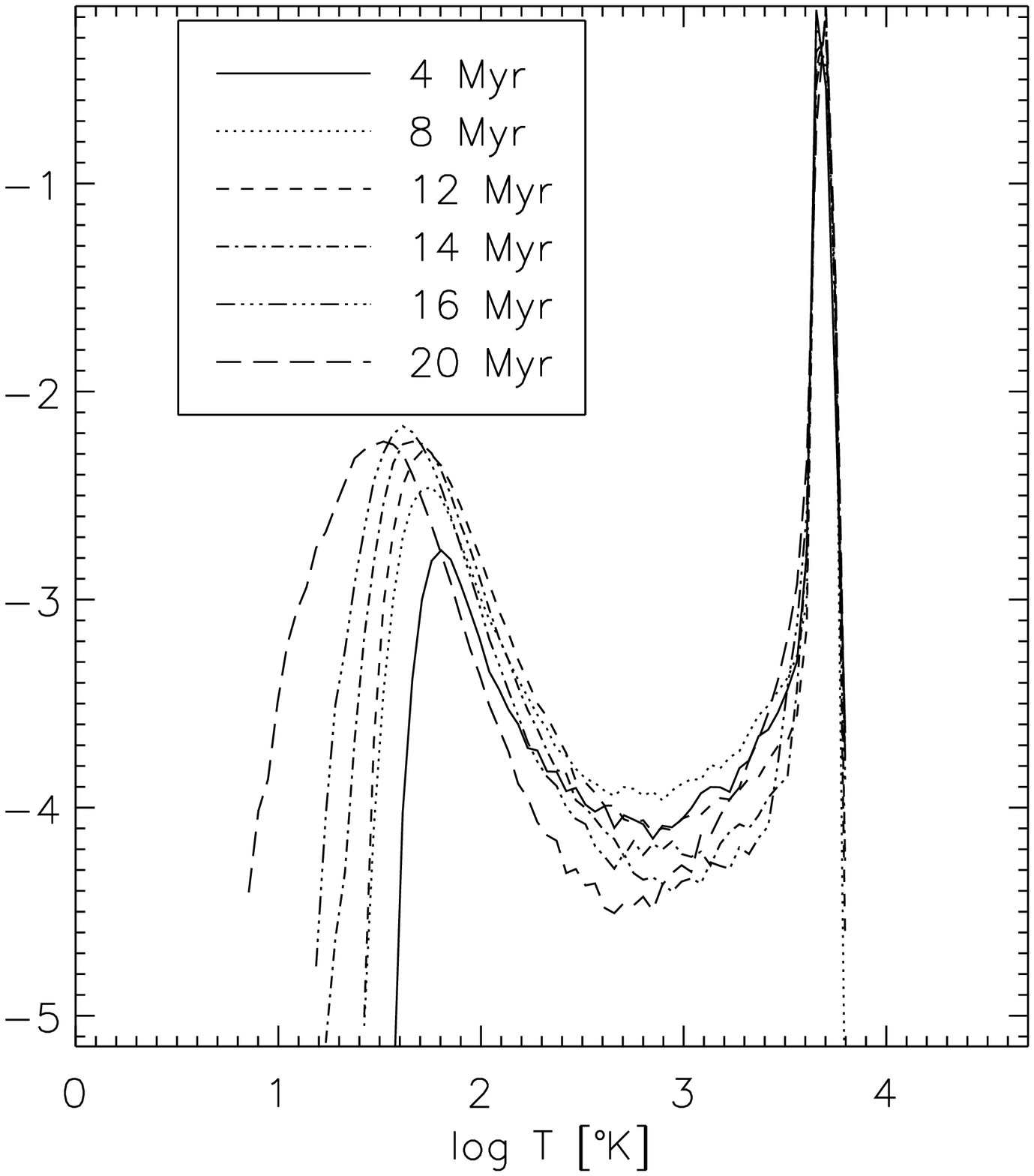}
\caption{Evolution of the mass-weighted density (\emph{left}
frame) and temperature (\emph{right} frame) histograms (normalized to
the maximum) for run
L256$\Delta v$0.17 at various times. \label{fig:hists} }
\end{figure}

An important consideration for understanding the production of
local collapse events is the distribution of the density and
temperature as the cloud evolves. This is shown in fig.\
\ref{fig:hists}, where the normalized mass-weighted histograms of density
(\emph{left} panel) and of temperature (\emph{right} panel) are
shown at times 4, 8, 12, 14, 16, and 20 Myr for run L256$\Delta v$0.17. It can
be seen that a cold phase already exists at $t=4$ Myr, although
at this time the cloud's conditions do not greatly exceed those
of the standard CNM, meaning that turbulence is only building up
in the cloud at this stage, at which the simulation resembles a
two-phase medium. At $t=8$ Myr, the density maxima and
temperature minima have shifted to more extreme values, which
persist up to $t=12$ Myr, indicating that the density and
temperature fluctuations are predominantly created by the
supersonic turbulence in the cloud \citep{PaperI}. However, by
$t=14$ Myr, the density and temperature extrema are seen to be
again moving towards more extreme values, and this trend persists
throughout the rest of the simulation, indicating that
gravitational collapse has taken over the evolution of the
density fluctuations (see also \S \ref{sec:energy_sf}). 

It thus
appears that in this particular simulation, the turbulent density
fluctuations act as simply as seeds for the subsequent growth of
the fluctuations by gravitational instability, as proposed by
\citet{CB05}. However, both in that paper as in the present
study, the global turbulence is already decaying, and by
definition it cannot then continue competing with self-gravity.
Minimally, a fundamental role of turbulence, even in this decaying
state, must be to create \emph{nonlinear} density fluctuations, of much
larger amplitudes than would be created by thermal instability alone,
which can collapse in shorter times than the whole cloud (cf.\ \S\S
\ref{sec:energy_sf} and \ref{sec:implications}). In any case,
simulations in which the inflow lasts beyond the onset of
collapse would be desirable, but unfortunately, as mentioned in
\S \ref{sec:eff_infl_sigma}, we have not found it feasible to
attempt them with the present code and computational resources,
and such a study must await a different numerical scheme and
physical setup.


\subsection{Energy evolution and star formation} \label{sec:energy_sf}

\begin{figure}
\plotone{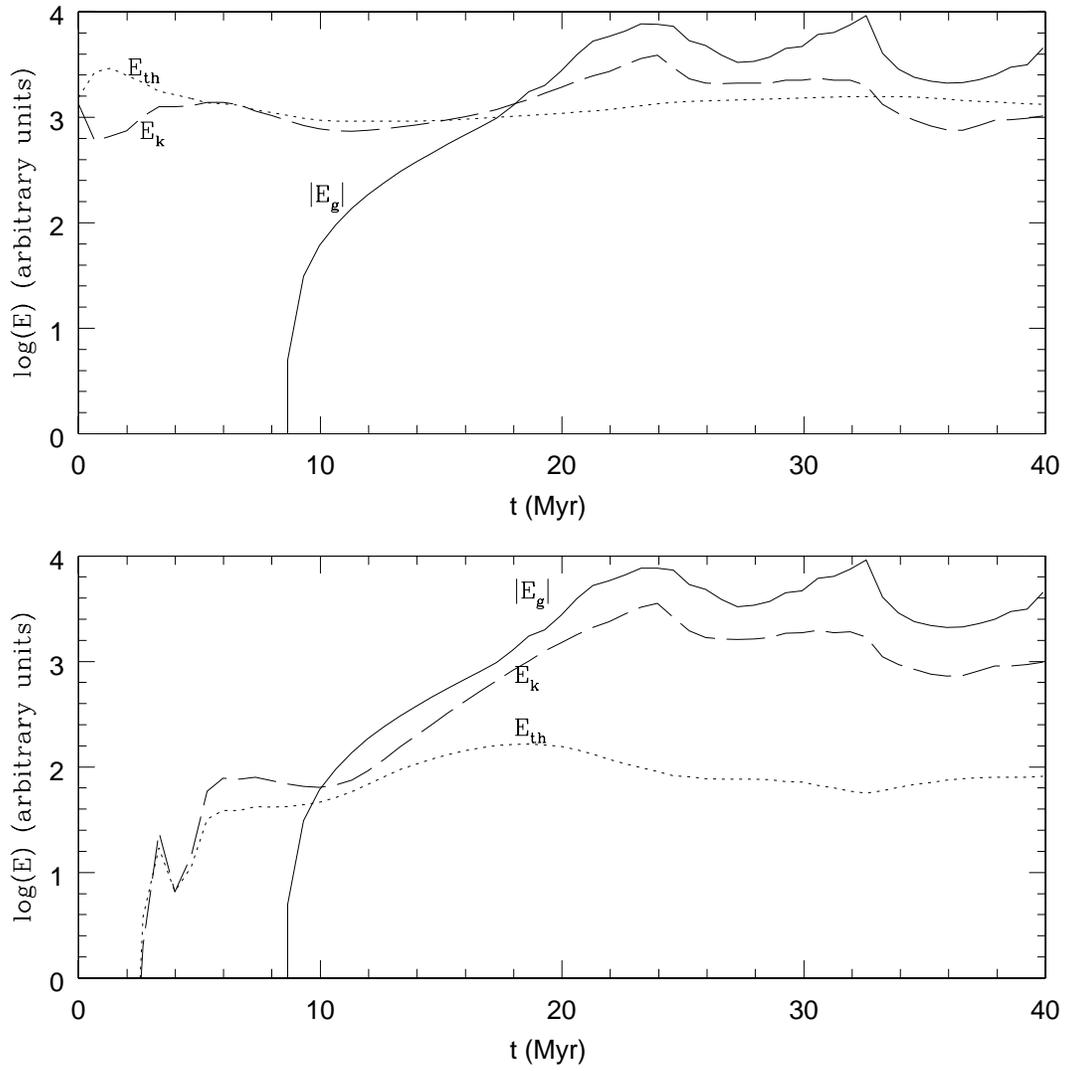}
\caption{\emph{Top}: Evolution of the total gravitational energy (in absolute
value) of the numerical box and of the kinetic and thermal energies in a
cylinder of radius 32 pc and length 16 pc centered in the middle
of the numerical box for run L256$\Delta v$0.17. \emph{Bottom}: Similar to the
top panel, but with the
thermal and kinetic energies calculated for the dense gas ($n > 50 \pcc$)
only.
\label{fig:multi_evo2}
}
\end{figure}

Figure \ref{fig:multi_evo2} (\emph{top} panel) shows the evolution of
the kinetic ($\Ek$) and thermal ($\Eth$) energies within a cylinder of
length 16 pc and radius 32 pc centered at the middle point of the
numerical box, together with the evolution of the (absolute value of
the) gravitational energy ($|\Eg|$) for the entire simulation
box.  {This}
cylinder contains 
 {most of the cloud's mass thoughout the simulation},
although  {it}
also contains sizable amounts of interspersed WNM. Figure
\ref{fig:multi_evo2} (\emph{bottom} panel), on the other hand, shows the
evolution of the same energies but with $\Ek$ and $\Eth$ calculated for
the dense gas only.

The gravitational energy is shown for the entire box because of the
practical difficulty to evaluate it only for the cloud's mass, although
we expect the latter to dominate the global gravitational energy when
the cloud has become very massive. Indeed, we observe that $|\Eg|$ at
late times is very large and dominates the other forms of energy. A
brief period of positive values of $|\Eg|$ is observed for $0 \lesssim t
\lesssim 9$ Myr, which can be understood as follows: Because the
boundary conditions are periodic, Poisson's equation for the
gravitational potential is actually modified to have the density
fluctuation $\rho - \langle\rho\rangle$ as its source. This is
standard fare in cosmological simulations
\citep[see, e.g., ][]{AL88, KT90}. Moreover,
gravity is a long-range force, and so large-scale features tend to
dominate $\Eg$. Thus, during the early epochs of the simulation, when
the cloud is beginning to be assembled, the dominant density features
are the voids left behind by the cylinders because they are
large, while the cloud is small and not very massive. However, once the
cloud becomes sufficiently massive, and the voids have been smoothed
out, the cloud dominates $\Eg$. {The period of positive
$\Eg$ is omitted from both panels of fig.\ \ref{fig:multi_evo2}}.

\begin{figure}
\epsscale{1.}
\plotone{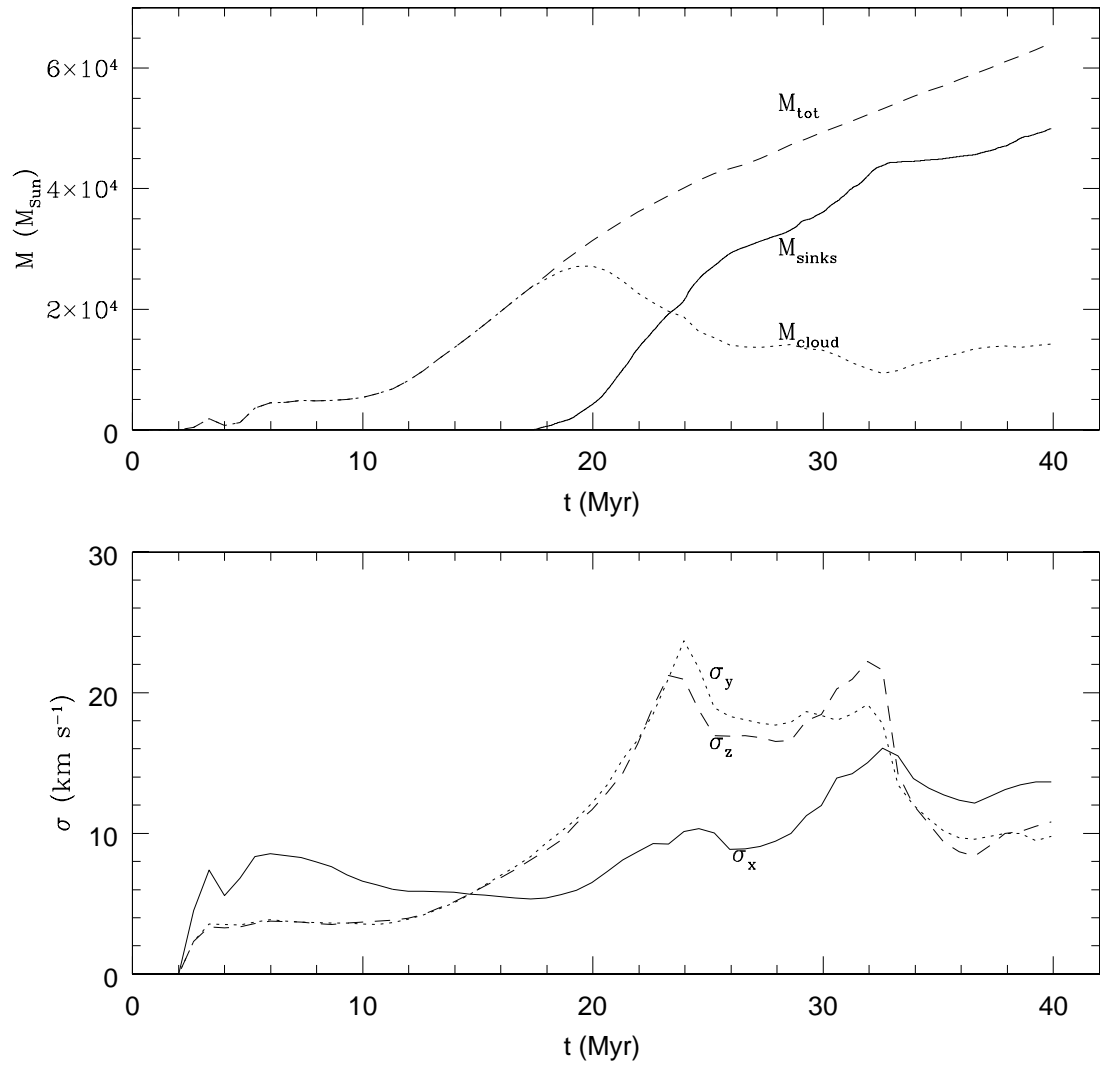}
\caption{\emph{Top}: Evolution of the cloud, {sink}, and
total {(cloud+sink)} mass for
run L256$\Delta v$0.17. \emph{Bottom}: Evolution of the velocity dispersion of 
the dense gas ($n > 50 \pcc$) in each of the three coordinate
axes.
{From $t \approx 6$ Myr through $\approx 12$ Myr, $\sigma_x$
decreases, reflecting the smooth end of the inflows.
At this time, the gravitational collapse of the cloud (in the $yz$
plane) sets in, generating an increase in $\sigma_y$ and $\sigma_z$.
While initially unaffected by this collapse, $\sigma_x$ increases
starting at $t \approx 18$ Myr, coinciding with the onset of star
formation.
}
\label{fig:multi_evo} 
} 
\end{figure}

Comparison of the energy plots (figs.\ \ref{fig:multi_evo2} \emph{top} and
\emph{bottom}) with the evolution of the gas and sink masses (fig.\
\ref{fig:multi_evo} \emph{top}) and the animations shows some very
important points. First and foremost, the cloud is never in virial
equilibrium over its entire evolution. Instead, we can identify
three main stages of evolution. First, over the period
$8.5 \lesssim t \lesssim 14$ Myr, $|\Eg|$ {increases}
monotonically, transiting from being negligible compared to
$\Ek$ and $\Eth$ to becoming larger than either one of them. The
exact time at which this occurs depends on 
what system is being considered. It occurs at $t \sim 10$ Myr
when only the energies in the dense gas are considered, while it
occurs at $t \sim 18$ Myr when the entire cylinder, which
includes substantial amounts of warm gas, is considered. Since
global gravitational contraction starts at $t \sim 12$ Myr,
it appears that the true balance is bracketed by the estimates
based on the full cylinder volume and on the dense gas, though
closer to the latter. Thus, over the interval $8.5 \lesssim t
\lesssim 12$ Myr, the increase in $|\Eg|$ is driven mostly by the
cooling and compression of the gas.

Second, from $t \sim 12$ Myr to $t\sim 24$ Myr, $|\Eg|$ continues
to increase ($\Eg$ becomes more negative), but now driven by the
global collapse of the cloud. Over this period, we see that $\Ek$
for the dense gas (fig.\ \ref{fig:multi_evo2}, \emph{bottom}
panel) closely follows $|\Eg|$, approximately satisfying the
condition 
\begin{equation}
|\Eg| = 2 \Ek.
\label{eq:equip}
\end{equation}
We see that in this case, \emph{this condition is a signature of
collapse, not of gravitational equilibrium}, even though it lasts
for nearly 12 Myr. The maintenance of this equipartition arises
from the fact that the cloud is converting gravitational
potential energy into kinetic energy, but is replenishing the
former by the collapse itself. 

Note that we have cautioned in \S
\ref{sec:cloud_disrup} that our simulations may not be realistic after
$t \approx 20.3$ Myr, because by that time the stellar energy feedback
may be sufficient to revert the global collapse and disperse the
cloud, but even in that case the cloud will
have evolved  {out of}
equilibrium up to that point. 

Finally, after $t \sim 24$ Myr, the gravitational and kinetic
turbulent energies saturate and begin to vary in approximate
synchronicity. Although this late stage may not be representative
of actual MCs, it is important to understand what is happening
in the simulation. The near constancy of $|\Eg|$ and $\Ek$ could
naively be interpreted as a final state of near virial
equilibrium. However, inspection of the animations shows
otherwise. At $t \sim 24$ Myr, the main body of the cloud is
completing its global collapse, and the mass in stars is
beginning to exceed the mass in dense gas, which is itself
decreasing. So, at this point, the gravitational energy is
beginning to be dominated by the stars, rather than by the
gas. Moreover, the face-on animation (fig.\
\ref{fig:anim_256_faceon}) shows that at this time
($t=180$ in code units, shown in the animations) the star cluster
is beginning to re-expand, after having reached maximum
compression. This leads to the decrease of $|\Eg|$ between $t
\sim 24$ and $t \sim 28$ Myr. Meanwhile, $\Ek$, which is
dominated by the dense gas, decreases because the dense gas
is being exhausted. Nevertheless, the outer ``chaff'' of the
dense gas, mostly in the form of radial filaments, is continuing
to fall onto the collapsed cloud. This leads to a new increase in
both energies, as this is a secondary collapse. This situation
repeats itself at $t \sim 32$ Myr ($t=244$ in code units) after
the secondary collapse ends and its second-generation star
cluster begins to expand away. The end of each collapse and the
re-expansion of the clusters is marked by kinks in the graph of
$M_{\rm sinks}$ \emph{versus} time in fig.\ \ref{fig:multi_evo}
(\emph{top}), indicating a decrease in the star formation rate.

It is important to recall as well that the condition given by
eq.\ (\ref{eq:equip}) is not sufficient for virial
equilibrium. The necessary and sufficient condition for this is
that the second time derivative of the moment of inertia of the
cloud be zero, and eq.\ (\ref{eq:equip}) cannot guarantee
this, as many other terms enter the full virial balance of the
cloud \citep{BVS99, Ball06, Dib_etal06}. Thus, the observed closeness of
$|\Eg|$ and $\Ek$ in actual molecular clouds
must exclusively be considered an 
indication of near equipartition and probably of collapse, but
not of virial equilibrium \citep[see also][]{KBVD05}.

Thus, rather than evolving in near virial equilibrium, \emph{the cloud
evolves far from equilibrium}. $|\Eg|$ starts out from essentially zero and
increases monotonically until it catches up with the thermal and
kinetic energies. From that time on, gravity dominates the energy
balance, leading to collapse, which in turn causes a near
equipartition between $|\Eg|$ and $\Ek$, although both energies
continue to vary systematically.

Finally, figure \ref{fig:multi_evo} (\emph{top}) also shows that 
star formation begins at $t \sim 17.2$ Myr, roughly 5 Myr after
global contraction of the cloud has begun. Yet, the star-forming
local collapse events proceed much more rapidly than the global
collapse of the cloud. This indicates that
the turbulence in the cloud has created \emph{nonlinear} density
fluctuations, whose local free-fall time is shorter than that of
the whole cloud. In particular, the mass in stars increases from
zero to $\sim 15$\% of the cloud's mass in $\sim 3$ Myr (from $t=17$ to
$t=20$ Myr).

The main conclusions from this section are that a) the cloud evolves
far from equilibrium all the way from its inception through its
final collapse; b) as soon as $\Eg$ dominates the dynamics, the
cloud develops equipartition indicative of the collapse, not of
equilibrium, and c) star formation is rapid
in comparison with the evolution and collapse of the whole cloud.


\subsection{Effect of the inflow duration and initial velocity
dispersion} \label{sec:eff_infl_sigma}

In previous papers \citep{VKB05, PaperI}, it has been
argued that the star formation efficiency is a sensitive
function of whether the turbulence in the cloud is in a driven or
in a decaying regime. Our system is driven at
first, and gradually transits to a decaying regime, as pointed out in \S
\ref{sec:turb}.  In all of
the runs in this paper, global cloud contraction and the
subsequent star formation occur after the inflows have weakened
substantially ($t > \ti$). The main motivation behind run
L256$\Delta v$0.17 was precisely to model an inflow of as long a duration
as possible, to approximate the case of a driven cloud, although
this goal was not completely achieved. For example, \citet{HBB01}
suggest accumulation lengths of up to 400 pc. Runs with even larger
boxes (e.g., 512 pc) would be desirable, but they are either prohibitely
expensive, or have an excessively poor mass resolution. In addition,
vertical stratification effects would have to be
considered. This will require a transition to the code Gadget2,
which allows for non-cubic boundary conditions and/or inflow
boundary conditions, a task we defer to a subsequent paper.

Nevertheless, comparison of the runs with different inflow
lengths does shed light on the effect of a longer driving
duration. The cloud in run L256$\Delta v$0.17, whose inflow has $\li = 112$
pc and lasts $\sim 22$ Myr (although it begins weakening at $T \sim \ti
= 12.2$ Myr), begins contracting at $t \sim 12.5$ Myr and begins
forming stars at $t \sim 17$ Myr, while both runs with $\li = 48$
pc, for which the inflow lasts 11 Myr (begins weakening at $t \sim
\ti = 5.2$ Myr), start to contract at $t \sim 8$ Myr (recall that
the inflows have the same density and 
velocity in all runs). Run L128$\Delta v$0.24 begins forming stars at $t =
15.75$ Myr, and run L128$\Delta v$0.66 does so at $t = 13.77$ Myr. 

\begin{figure}
\plotone{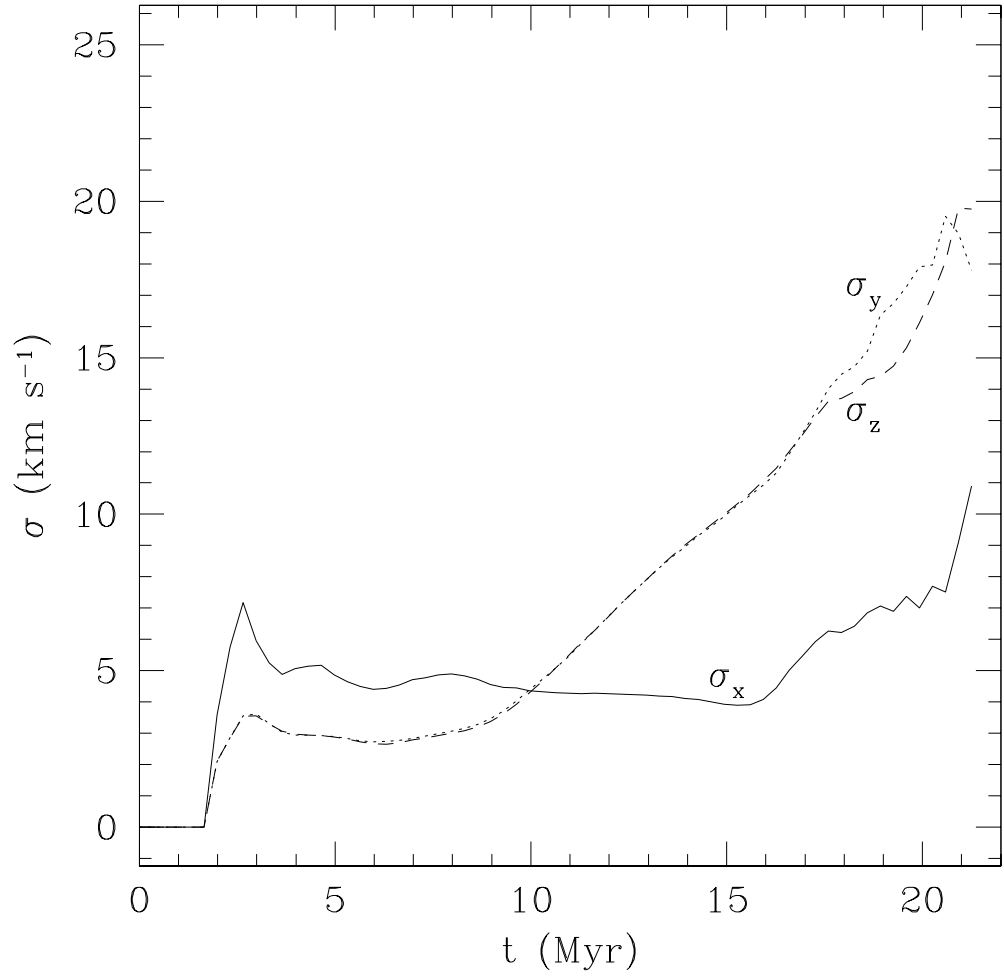}
\caption{Evolution of the velocity dispersion in the dense gas ($n >
50 \pcc$) along each of the three coordinate axes for run
L128$\Delta v$0.24. Compare to fig.\ \ref{fig:multi_evo}
(\emph{right}), noting the different extension of the
axes. \label{fig:veldisp_run18}
}
\end{figure}

These results
clearly suggest that a longer inflow duration delays the onset
of both global collapse and star formation, in spite of the fact
that the cloud formed by it is more massive. This is attributable
to the fact that the turbulence in the cloud is continuously
driven by the inflow. This is verified by comparing the three
components of the velocity dispersion for runs L256$\Delta v$0.17 and
L128$\Delta v$0.24, respectively shown in fig.\
\ref{fig:multi_evo} (\emph{bottom}) and fig.\ 
\ref{fig:veldisp_run18}. Both runs show an initial
transient peak in $\sigma_x$ ending at $t \approx 4$ Myr. However, after
this transient, $\sigma_x$ in run L256$\Delta v$0.17 increases again, reaching a
maximum of $\sigma_x \sim 8.5 \kms$, and decreasing afterwards, until it
nearly stabilizes at a value $\sigma_x \sim 6 \kms$ at $t \sim 11.5$
Myr. During this time interval, 
$\sigma_y \approx \sigma_z \sim 4 \kms$. Instead, in run
L128$\Delta v$0.24, $\sigma_x$ does not increase again after the initial
transient, and remains at a much more moderate level of $\sigma_x \sim 5
\kms$, while $\sigma_y \approx \sigma_z \sim 3 \kms$. Clearly, the
turbulence level is significantly higher in the longer-inflow run.

Note that the delay in the onset of global collapse and of local star
formation cannot be attributed to the amplitude of the
initial velocity fluctuations in the inflow, since runs L128$\Delta v$0.24 and
L256$\Delta v$0.17, which have comparable amplitudes, differ substantially in
these times, while runs L128$\Delta v$0.24 and L128$\Delta v$0.66 begin global
contraction at almost the same time and differ by only $\sim 15$\% in
the time at which they begin forming stars, in spite of one
having more than twice the velocity fluctuation amplitude of the
other. This leaves the inflow duration as the sole cause of delay of
both the global and local collapses.


\subsection{Local column density and star formation} \label{sec:N_SF}

The initial conditions in our simulations are assumed to consist
exclusively of atomic gas. However, the final density and
temperature conditions are typical of molecular clouds, so
molecule formation must occur somewhere along the way in the
evolution of the clouds. One important shortcoming of our
simulations is that they cannot distinguish between atomic and
molecular gas, as no chemistry is included. Thus, we cannot
directly tell from the simulations how long does it take for star
formation to begin after molecular gas forms. We can,
nevertheless, measure the column density of the dense gas in
star-forming regions, and compare it to the estimates for
self-shielding given by \citet{FC86} and \citet{HBB01}, under
solar Galactocentric conditions of the background UV field. The
former authors find a threshold column desity for self-shielding
of $\Nss \sim 5 \times 10^{20} \psc$, while the latter authors
quote values $\Nss \sim$ 1--2$ \times 10^{21}\psc$ from \citet{vDB88} and
\citet{vDB98}. So, we take a reference value of $\Nss = 10^{21}
\psc$.

For comparison, in Table \ref{tab:coldens_SF} we report the
column densities of the dense gas in the first four regions of
star formation in run L256$\Delta v$0.17 at the time immediately before
they begin forming stars. We see that the
column densities fall in the range 0.5--2 $\times 10^{21}
\psc$, suggesting that star formation \emph{locally} occurs nearly
simultaneously with the conversion of the gas from atomic to molecular. 

Moreover, we can compare the time at which we expect the bulk of the
cloud to become molecular (i.e., the time at which the mean column
density in the cloud is reaching $\Nss$) with the time at which star
formation begins in the cloud. A lower limit for the time at which the
mean column density in the cloud equals $\Nss$ is given by
\begin{equation}
t_{\rm ss} \gtrsim \frac{\Nss}{2 n_0 \vi} = \frac{\Nss}{5.8 \times 10^{19}
\psc}~{\rm Myr},
\label{eq:mean_N}
\end{equation}
where $n_0$ and $\vi$ have been defined in \S \ref{sec:params}. This
estimate is a lower limit because, as mentioned in \S \ref{sec:mass},
$\vi$ is not constant, but actually decreases in time. For $\Nss =
10^{21} \psc$, we obtain $t_{\rm ss} \gtrsim 17.3$ Myr, suggesting that the
bulk of the cloud is still expected to be predominantly atomic by the time star
formation is beginning.

Together, these simple estimates suggest that stars form roughly
simultaneously with the conversion of gas from atomic to molecular, with
the local star forming regions being more advanced in the conversion
process than the bulk of the cloud. In turn, this suggests that our
model cloud would be observed as a collection of predominantly molecular clumps
immersed in a large, predominantly atomic, substrate. This suggestion is
consistent with recent observations that substantial amounts of atomic
gas coexist with the molecular phase in MCs \citep{LG03, GL05,
Klaassen_etal05}, with perhaps even warm gas mixed in \citep{HI06}.
Of course, a more conclusive
confirmation of these estimates must await simulations in which the
chemistry and radiative transfer are properly taken care of.


\section{Discussion} \label{sec:discussion}

\subsection{Implications} \label{sec:implications}

\subsubsection{Dynamic evolution and global \emph{versus} local
collapse} \label{sec:glob_vs_loc}

The results presented in \S \ref{sec:results} provide general support
to the scenario outlined in \citet{Elm93b, BVS99, BHV99, HBB01, VBK03,
Hart03, MK04, Hetal05, Hetal06, PaperI, BKMV06}. The clouds in our
simulations are 
never in a state of virial equilibrium before they convert most of
their mass into stars. Instead, they are in a continuously evolving
state, initially obtaining their mass and turbulence
simultaneously as they form 
out of the compression and cooling of diffuse warm gas \citep[see
also][]{HP99, HP00, KI02, IK04}. The gas initially has negligible
self-gravity compared to its thermal support, but it quickly becomes
super-Jeans because of the compression, the cooling, and the mass
increase of the dense gas, until its self-gravitating energy
eventually becomes comparable with the sum of its thermal and
turbulent energies, at which point it begins to undergo global
collapse. After this time, gravity becomes the main driver of the
large-scale motions in these ({semi-}) decaying simulations, 
causing a near-equipartition, $|\Eg| \sim 2 \Ek$, which however
is indicative of \emph{collapse}, not equilibrium. 

The nonlinear density fluctuations induced by the turbulence
collapse faster than the whole cloud, as they have shorter
free-fall times, and star formation proceeds vigorously before
global collapse is completed. This result is in contrast with the
standard notion that
\emph{linear} density fluctuations cannot lead to fragmentation
because the fastest growing mode of gravitational instability in a
nearly uniform medium is an overall contraction of the whole medium
\citep{Larson85}. Thus, a crucial role of the turbulence in the medium
is to create \emph{nonlinear} density fluctuations that have shorter
free-fall times than the entire cloud.

These results bring back the scenario of global cloud collapse
proposed by \citet{GK74}, but with a twist that avoids the
criticism of \citet{ZP74}, namely that MCs could not be in global
collapse because the star formation rate would be exceedingly
high. Our simulations suggest that MCs may be undergoing global
gravitational contraction, but the efficiency is reduced
because local collapse events, which involve only a fraction of
the total cloud mass, proceed faster than the global
collapse. {Once a sufficiently large number of stars have
formed, their energy feedback may partially or completely halt the
collapse.} Further reduction may be provided by supercritical magnetic fields
\citep{VKB05,NL05}. This scenario is consistent with the recent proposal of
\citet{HB06} that the Orion A cloud may be undergoing
gravitational collapse on large scales.

\subsubsection{Cloud's mass variation and cloud boundaries}
\label{sec:mass_var_vs_bound} 

The fact that the cloud's mass is not constant is equivalent to the
property that the locus of a Lagrangian boundary of the cloud defined
(by means of a threshold density) at any given time does not remain at
the cloud's boundary as time progresses, but instead it is later
incorporated into the interior of the cloud. This implies that the cloud
does not always consist of the same set of fluid parcels, and that in a
virial balance analysis of the cloud, there is non-zero flux of the
physical variables through Eulerian boundaries defined at any given time
\citep{BVS99, SVB02, Dib_etal06, Ball06}.

\subsubsection{Absence of slow, accelerating star formation
phase} \label{sec:no_accel}

It is important to note that the star formation rate (given
by the slope of the mass in sinks \emph{versus} time in fig.\
\ref{fig:multi_evo} [\emph{top}]) is large from the start, and we
observe no long period of slow, accelerating star formation, contrary to
the suggestion of \citet{PS00, PS02} that star formation accelerates in
time. Problems with this suggestion have been discussed by \citet{Hart03},
and, within the framework of their assumptions and limitations, our
simulations do not confirm it. {An important question is
whether this result will persist when magnetic fields are included. It
is possible that during the early stages of the cloud's evolution, it
may behave under magnetically subcritical conditions, giving very low
star formation rates (SFRs), and then transit into a supercritical regime, with
higher SFRs (\S \ref{sec:accum_dist}).}

\subsubsection{Latency period} \label{sec:latency}

The clouds in our simulations do spend a long latency period
between the beginning of the compressive motion that creates the
cloud and the time at which they begin forming stars ($\sim$
14--17 Myr in the three runs we have considered).  However, this
long period is most probably spent in the atomic phase, since the
column densities of the star-forming regions in the simulations
are comparable to those required for molecule
formation \citep{FC86, HBB01}. {More specifically, \mbox{\cite{Bergin_etal04}} have found that the
timescale for CO molecule formation is 
essentially that required for reaching a dust extinction of $A_V \sim
0.7$. Note also that during this
time, the formation of H$_2$ can be fast within density peaks
(cores), even if most of the mass of the
cloud is still in the low-density regime \citep{PAV02, GM06}. Thus, the
concept of ``rapid'' 
star formation refers to the time elapsed between the appearance of a
(CO) molecular cloud and the onset star formation, as well as to the
\emph{rate} of the star formation process itself.}

Note that the latency period is in fact by definition of the order of
the crossing time of the entire {large-scale} compressive
wave, in agreement with observational evidence
\citep{Elm00}. 

\subsubsection{Implications for HI envelopes and magnetic criticality}
\label{sec:accum_dist}

{The formation of GMcs by an accumulation process such as
the one modeled here has two important additional implications. First,
because the process starts with lower-density gas (the WNM in our
simulations) that is compressed by some external agent, GMCs, which are
the ``tip of the iceberg'' of the density distribution, are expected to
have in general HI envelopes, which would constitute the corresponding
``body of the iceberg''. This gas would include transition material
traversing the unstable phase between the warm to the cold
phases, which would appear to have a nearly isobaric behavior
\citep{GVK05, AH05, PaperI}, in agreement with the conclusion
{reached} by
\citet{AW93} {from observations of HI envelopes of MCs}. Such
envelopes are routinely observed 
\citep[e.g., ][ see also the combined HI and CO maps of Blitz \&
Rosolowsky 2004]{BT80, WLM83, AWM91, AW93, BHV99, Brunt03,
Blitz_etal06}. Weaker compressions than we have modeled
may produce mostly thin CNM sheets, with little or no molecular gas, as
discussed in \citet{PaperI}.

Second, this scenario of molecular cloud formation implies that the
mass-to-flux ratio of the cloud is a variable quantity as the cloud
evolves. This ratio is equivalent to the ratio of column density to
magnetic field strength \citep{NN78}, with the critical colum density
given by $\Sigma \sim 1.5 \times 10^{21} \left[B/5\mu G \right]~\psc$.
Although in principle under ideal MHD conditions the criticality of a
magnetic flux tube involves \emph{all} of the mass contained within it,
in practice it is only the mass in the dense gas phase that matters,
because the diffuse gas is not significantly self-gravitating at the
size scales of MC complexes. As pointed out by \cite{HBB01}, the above
value of the dense gas' column density for magnetic criticality is very
close to that required for molecule formation (eq.\
[\ref{eq:N_self_shield}]) and for gravitational binding (eq.\
[\ref{eq:N_grav_unst}]), and therefore, the cloud is expected to become
magnetically supercritical nearly at the same time it is becoming
molecular and self-gravitating. }


\subsection{Limitations and future work} \label{sec:caveats}

Our simulations are limited in a number of aspects. First and
foremost, as already mentioned in \S \ref{sec:sinks}, one major
shortcoming of our simulations is the absence of feedback from the
stellar objects onto the cloud. This is certainly an unrealistic
feature. Second, our clouds are in a regime of turbulence decay after
the inflows have subsided. This may or may not be an unrealistic
feature, and rather it {may}
represent a fraction of the
population of clouds. Finally, we have neglected magnetic fields
altogether, due to the non-existence of suitable SPH algorithms
including MHD for the problem of fully developed turbulent flows.

All of these limitations tend to exaggerate the SFE in our
simulations, as the turbulence within them is not replenished by
either a continuing flow or by stellar energy feedback. It is possible
that a continued inflow (and therefore a sustained turbulence level)
could prevent global collapse altogether, with only local collapse
events happening randomly, although certainly the inflows cannot last
indefinitely. Thus, it would appear that global collapse is
inescapable. However, once star formation begins, the stellar energy
feedback is likely to either be able to halt the global collapse and disperse
the cloud \citep{FST94, HBB01} {or else
maintain it in rough equilibrium {\citep[e.g., ][]{Matz02, NL05,
TKM06, KMM06}}}. 
Thus, {two} important
questions to address in future papers {are, one,}  
what is the evolution like when star formation begins before the
inflows subside. It is 
likely that this case will have much smaller SFEs. {Two,
whether the stellar energy feedback tends 
to disperse the clouds, or else to
maintain them in rough 
equilibrium. Observationally, star clusters older than 5--10 Myr tend to
be already devoid of gas {{\citep{LBT89, HBB01, BH06}}},
suggesting that the effect is more disruptive than equilibrating.} 

Finally, our simulations {have} neglected the magnetic field. As discussed in
\S \ref{sec:quali}, this is not a crucial omission if GMCs are in
general magnetically supercritical, since the supercritical case is
qualitatively equivalent to the nonmagnetic case, with the only
difference being that it behaves as
if it were less massive than a non-magnetic cloud of the same mass
\citep{Shu92, Hart98}. Thus, we consider that our models are still
representative of the large-scale evolution up to the early times of
star formation in supercritical clouds. The main difference is
expected to be that magnetized, supercritical clouds should have lower
SFEs than nonmagnetic ones \citep{VKB05, NL05}. On the other hand, the
evolution of subcritical clouds will certainly differ from the models
presented here, because they will have no global tendency to collapse.
At any rate, it is necessary to perform simulations of the full
process in the presence of magnetic fields. This will presumably
require the usage of AMR techniques incorporating the
analogue of sink 
particles and stellar feedback, and will be pursued in future papers.


\section{Summary and conclusions} \label{sec:concl}

In this paper we have presented a suite of numerical simulations
designed to investigate jointly the formation of molecular clouds (MCs)
and of stars within them. The simulations use an SPH scheme including
self-gravity, sink particles and cooling {leading to
thermal bistability}. Magnetic fields are neglected.
The simulations describe the collision of
oppositely-directed gas streams (``inflows'') in the warm neutral medium
(WNM) at moderately supersonic velocities (each with a velocity of $9.2
\kms$, or a Mach number of 1.22 in the unperturbed WNM). Three
simulations were considered, varying the length of the inflow and the
amplitude of the initial velocity fluctuations in the gas. The
collisions trigger a transition to the cold phase in the gas, and
simultaneously generate turbulence in the resulting
``cloud'', defined as 
the gas at densities $n > 50 \pcc$. The inflows
secularly weaken in time, and so does the turbulence level in the
cloud, implying that {the turbulence in the
clouds gradually transits from being continuously driven to being in a
decaying regime}. The cooling and the mass gain of the dense 
gas eventually cause the cloud to contain a large number of Jeans masses
at the mean conditions. Moreover, because the cloud is supersonically
turbulent, locally the density can be much larger, with
a correspondingly lower Jeans mass and shorter free-fall times than
{those} of the whole cloud. 

Thus, by the time the inflows have almost subsided, the cloud engages
in global gravitational collapse, but shortly thereafter it begins to
produce numerous local collapse events that occur on much smaller
timescales because of the larger densities, so that by the time the
global collapse of the cloud is completed, it has converted most of
its mass to collapsed objects (sink particles). The sinks in general
represent star clusters, due to the limited mass resolution and lack of
modeling of opacity-limited fragmentation \citep[e.g.][]{BBB03}.
Our simulations can only be considered reliable up to the time
when the mass in collapsed objects (sinks) implies a high enough
number of massive stars that they would disrupt the cloud. Nevertheless, the
evolution up to that point shows a number of relevant results:

\medskip
\noindent
1. The clouds are never in virial equilibrium during this
period. Instead, they continually evolve, increasing their mass and
self-gravitating energy, until the latter becomes comparable or larger
than the thermal and kinetic energies. {Some} $\sim 5$ Myr {after global
contraction began}, star
formation begins.

\medskip
\noindent
2. In spite of not being in equilibrium, the near-equipartition
condition $|\Eg| \sim 2 \Ek$ is approximately satisfied as soon
as $|\Eg|$ becomes comparable to $\Ek$, because of the
gravitational contraction of the cloud, with both quantities
increasing simultaneously. This occurs long before the onset of
star formation. This fact can explain the observed state of
apparent virialization of GMCs.

\medskip
\noindent
3. The near equipartition is a signature of global gravitational
collapse, not equilibrium, and suggests a return to 
Goldreich \& Kwan's (1974) scenario of global gravitational contraction
in MCs.  {However, the criticism by
\citet{ZP74}, namely that an excessively large star formation should result
through this process, is avoided in part because the nonlinear turbulent
density fluctuations collapse earlier than the whole cloud, involving
only a fraction of the total mass, and in part because as soon as the
stars form they probably contribute to dispersing the cloud, or at least
halting its global collapse. Further reduction of the SFE may occur
in the presence of supercritical magnetic fields.} 

\medskip
\noindent
4. Local collapse events begin to occur after global collapse has
begun, but they occur rapidly, requiring only $\sim 3$ Myr to convert
$\sim 15$\% of the cloud's mass into stars.

\medskip
\noindent
5. The star formation rate, measured by the slope of the mass in
stars \emph{versus} time, is large from the beginning. Within the framework,
assumptions and limitations of our simulations, no {long}
period of slow, 
accelerating star formation is observed.


\medskip
\noindent
6. Longer inflow durations maintain larger turbulent velocity
dispersions in the clouds and delay the onset of both global and local
collapse. Instead, larger amplitudes of the initial velocity
fluctuations have little effect in delaying the collapse. The latter
effect can be attributed to the fact that the turbulence is already
dissipating in the clouds by the time global collapse begins.

\medskip
\noindent
7. A long period ($\sim 14$--17 Myr) of ``dormancy'' does occur
{between the
time when the cloud begins to form and the time when} star formation
begins. Nevertheless, it is likely that 
most of this time is spent with the gas being in atomic form, since the column
density of star forming regions in our simulations are comparable to
values required for molecular gas formation, as reported in the
literature. Thus, our simulations support the notion that star formation
occurs almost simultaneously with the formation of molecular gas, under
local ISM conditions.

\medskip
We conclude that our simulations support the scenario of rapid star
formation after molecular gas has formed, involving accumulations of gas
from distances of a few hundred parsecs \citep{HBB01, VBK03}, while
simultaneously requiring timescales of the order of the crossing time
across the largest scales involved \citep{Elm00}, and in a
systematically out-of-equilibrium fashion \citep{BVS99, BHV99, KHM00}. However,
the evidence will have to be made more compelling as additional physical
processes, such as longer-duration inflows, stellar feedback, magnetic
fields and chemistry are included.

\acknowledgements
We thankfully acknowledge useful comments from Lee
Hartmann and Chris McKee, and the help of Luis
Ballesteros-Paredes with postprocessing of the computer animations. This
work has received financial support from CRyA-UNAM; from CONACYT grants
36571-E and 47366-F to E. V.-S., and UNAM-PAPIIT grant 110606; and from
the Emmy Noether Program of the German Science Foundation (DFG) under
grant Kl1358/1 to R.S.K.\ and A.K.J. The simulations were performed in
the cluster at CRyA-UNAM acquired with grant 36571-E. This work has made
extensive use of NASA's ADS and LANL's astro-ph abstract services.

\clearpage

\begin{deluxetable}{ccccccccccc}
\tablewidth{0pt}
\tablecaption{Run parameters. \label{tab:run_params}
}
\tablehead{
\colhead{Run name} & \colhead{$\Lbox$\tablenotemark{a} (pc)} &
\colhead{$\li$\tablenotemark{b} (pc)} &
\colhead{$\vi$\tablenotemark{c} (km s$^{-1}$)} &
\colhead{$\Mi$\tablenotemark{d}} & 
\colhead{$\vrmsi$\tablenotemark{e} (km s$^{-1}$)} &
\colhead{$M_{\rm box}$\tablenotemark{f} ($M_\sun$)} & 
\colhead{$\Npart$\tablenotemark{g}} & \colhead{$\Delta
M$\tablenotemark{h} ($M_\sun$)}\\  
}
\startdata
L128$\Delta v$0.24 & 128 & 48 & 9.20 & 1.22 & 0.24 & $6.582 \times 10^4$ &
$1.643 \times 10^6$ & 0.04 \\
L128$\Delta v$0.66 & 128 & 48 & 9.20 & 1.22 & 0.66 & $6.582 \times 10^4$ &
$1.643 \times 10^6$ & 0.04 \\
L256$\Delta v$0.17 & 256 & 112 & 9.20 & 1.22 & 0.17 & $5.253\times 10^5$ &
$3.242 \times 10^6$ & 0.16 \\

\enddata
\tablenotetext{a}{Physical size of computational domain.}
\tablenotetext{b}{Linear size of each inflow.}
\tablenotetext{c}{Speed of inflows.}
\tablenotetext{d}{Mach number of inflows, with respect to sound
speed of unperturbed WNM.}
\tablenotetext{e}{One-dimensional rms speed of initial imposed
velocity fluctuations.}
\tablenotetext{f}{Total mass in the computational domain.}
\tablenotetext{g}{Number of SPH particles.}
\tablenotetext{h}{Mass resolution, equal to the mass of each SPH particle.}
\end{deluxetable}

\begin{deluxetable}{ccccc}
\tablecaption{Column densities of star forming regions in run L256$\Delta v$0.17. \label{tab:coldens_SF}
}
\tablehead{
\colhead{Region} & \colhead{Time (Myr)} & \colhead{Position in $yz$
plane (pc)} & \colhead{$N$ ($\psc$)} & \\  
}
\startdata
1 & 17.0 & (144,135) & $9.58 \times 10^{20}$\\
2 & 17.26 & (131,142) & $2.27 \times 10^{21}$\\
3 & 18.06 & (112,125) & $4.89 \times 10^{20}$\\
4 & 18.59 & (137,115) & $1.52 \times 10^{21}$\\

\enddata
\end{deluxetable}

\clearpage

\end{document}